\documentclass[AMA,STIX1COL]{WileyNJD-v2}
\usepackage{graphicx,multirow,latexsym,mathtools,bbm}
\usepackage{float}
\articletype{Article Type}%

\newcommand{\bI}{{\bf I}}
\newcommand{\bH}{{\bf H}}
\newcommand{\bJ}{{\bf J}}
\newcommand{\bO}{{\bf O}}
\newcommand{\bS}{{\bf S}}

\newcommand{\bW}{{\bf W}}
\newcommand{\bX}{{\bf X}}

\newcommand{\bZ}{{\bf Z}}

\newcommand{\bh}{{\bf h}}

\newcommand{\br}{{\bf r}}

\newcommand{\bv}{{\bf v}}
\newcommand{\bx}{{\bf x}}
\newcommand{\by}{{\bf y}}

\newcommand{\sa}{\sigma}
\newcommand{\sat}{\sa^2}

\newcommand{\sate}{\sat_{\epsilon}}
\newcommand{\sata}{\sat_{\alpha}}

\newcommand{\bbeta}{{\mbox{\boldmath{$\beta$}}}}

\newcommand{\bth}{{\mbox{\boldmath{$\theta$}}}}
\newcommand{\bdelta}{{\mbox{\boldmath{$\delta$}}}}
\newcommand{\bme}{{\mbox{\boldmath{$\epsilon$}}}}

\newcommand{\bSigma}{\mbox{\boldmath{$\Sigma$}}}

\newcommand{\h}[1]{\hat{#1}}

\newcommand{\N}{\mathcal{N}}
\newcommand{\E}{\mathrm{E}}

\newcommand{\Cov}{\mathrm{Cov}}
\newcommand{\Var}{\mathrm{Var}}

\newcommand{\bzero}{\mathbf{0}}
\newcommand{\bone}{\mathbf{1}}

\newcommand{\bet}{\begin{theorem}}
\newcommand{\ent}{\end{theorem}}
\newcommand{\bep}{\begin{proposition}}
\newcommand{\enp}{\end{propposition}}
\newcommand{\bec}{\begin{corollary}}
\newcommand{\enc}{\end{corollary}}

\newcommand{\bel}{\begin{lemma}}
\newcommand{\enl}{\end{lemma}}

\newcommand{\beq}{\begin{eqnarray*}}
\newcommand{\enq}{\end{eqnarray*}}
\newcommand{\ben}{\begin{eqnarray}}
\newcommand{\enn}{\end{eqnarray}}
\newcommand{\bea}{\begin{equation}}
\newcommand{\ena}{\end{equation}}

\begin{document}

\title{Time-varying treatment effect models in stepped-wedge cluster-randomized trials with multiple interventions}

\author[1]{Zhe Chen}
\author[2]{Wei Wang}
\author[2]{Yingying Lu}
\author[2]{Scott D. Halpern}
\author[2]{Katherine R. Courtright}
\author[3,4]{Fan Li$^{\dagger,}$}
\author[1,2,5]{Michael O. Harhay$^{\dagger,}$}

\authormark{Z Chen \textsc{et al}}

\address[1]{\orgdiv{Department of Biostatistics, Epidemiology and Informatics}, \orgname{Perelman School of Medicine, University of Pennsylvania}, \orgaddress{\state{PA}, \country{USA}}}

\address[2]{\orgdiv{Clinical Trials Methods and Outcomes Lab, Palliative and Advanced Illness Research (PAIR) Center}, \orgname{Perelman School of Medicine, University of Pennsylvania}, \orgaddress{\state{PA}, \country{USA}}}

\address[3]{\orgdiv{Department of Biostatistics}, \orgname{Yale University School of Public Health}, \orgaddress{\state{CT}, \country{USA}}}

\address[4]{\orgdiv{Center for Methods in Implementation and Prevention Science}, \orgname{Yale University}, \orgaddress{\state{CT}, \country{USA}}}

\address[5]{MRC Clinical Trials Unit, University College London, London, UK}

\corres{Zhe Chen, Department of Biostatistics, Epidemiology and Informatics, Perelman School of Medicine, University of Pennsylvania, Philadelphia, PA 19104, USA.  \\
\email{zhe.chen@pennmedicine.upenn.edu}
 The $\dagger$ symbol denotes shared co-senior authorship.}

\abstract[]{The traditional model specification of stepped-wedge cluster-randomized trials assumes a homogeneous treatment effect across time while adjusting for fixed-time effects. However, when treatment effects vary over time, the constant effect estimator may be biased. In the general setting of stepped-wedge cluster-randomized trials with multiple interventions, we derive the expected value of the constant effect estimator when the true treatment effects depend on exposure time periods.
Applying this result to concurrent and factorial stepped wedge designs, we show that the estimator represents a weighted average of exposure-time-specific treatment effects, with weights that are not necessarily uniform across exposure periods. Extensive simulation studies reveal that ignoring time heterogeneity can result in biased estimates and poor coverage of the average treatment effect. 
In this study, we examine two models designed to accommodate multiple interventions with time-varying treatment effects:  (1) a time-varying fixed treatment effect model, which allows treatment effects to vary by exposure time but remain fixed for each time point, and (2) a random treatment effect model, where the time-varying treatment effects are modeled as random deviations from an overall mean. 
In the simulations considered in this study, concurrent designs generally achieve higher power than factorial designs under a time-varying fixed treatment effect model, though the differences are modest.
Finally, we apply the constant effect model and both time-varying treatment effect models to data from the Prognosticating Outcomes and Nudging Decisions in the Electronic Health Record (PONDER) trial. All three models indicate a lack of treatment effect for either intervention, though they differ in the precision of their estimates, likely due to variations in modeling assumptions.}

\keywords{stepped-wedge cluster-randomized trials, multi-arm randomized trial, time-varying treatment effect, linear mixed effects models, model misspecification}

\maketitle

\section{Introduction}

The stepped-wedge cluster-randomized trial (SW-CRT) is a pragmatic study design commonly used in public health and clinical research to evaluate the effect of interventions delivered at the group level \citep{gh,hh,overview}. Specifically, in this design, all individuals within a cluster or group initially receive the control condition (i.e., no intervention). Then, over time, clusters are sequentially randomized to start receiving the intervention at different steps or unidirectional crossover periods, and the study concludes when all clusters have transitioned to the intervention phase.

In standard analyses of stepped wedge cluster randomized trials, time effects are typically adjusted for as continuous or categorical covariates,\cite{zhang2023general} and the treatment effect is, due to the specification in the coding of the intervention as a 0/1 variable, assumed to be immediate and constant across time periods \citep{hh}. That is, once a cluster initiates the intervention, its treatment effect is parametrized to manifest instantly and remain unchanged over time—neither increasing nor decreasing.
In practice, however, this assumption may not hold, as the effect of an intervention may exhibit delays before becoming apparent or accumulate over time as clusters experience multiple periods of exposure to an intervention. When the constant treatment effect assumption is violated, estimates may become biased, and the model may fail to capture the true dynamics of the intervention over exposure time (i.e., the duration of the intervention), leading to unreliable conclusions. To this end, Kenny et al. \cite{kenny22} demonstrated that the expectation of the treatment effect estimator under a constant treatment effect model is a weighted sum of the true exposure time-specific treatment effects across exposure periods. However, this weighted sum (with possibly negative weights) does not necessarily approximate the average treatment effect, leading to potential bias. Moreover, the coverage of the associated confidence interval often falls below the nominal level, further compromising the validity of the inference. More recently, Wang et al.\cite{wang2024achieve} have also emphasized that, to achieve valid inference in standard stepped wedge designs using linear mixed models or generalized estimating equations, it is essential to correctly specify the treatment effect structure (e.g., whether it is constant or exposure time-dependent); misspecification of the other model aspects can usually be addressed by the use of a cluster-robust sandwich variance estimator.

The issue of heterogeneous treatment effects—where the intervention effect varies by exposure time or cumulative time on intervention—has received increasing attentions in the recent literature \citep{hgh15, nickless18, kenny22, maleyeff23, lee24, hughes24}. These approaches typically follow one of two strategies: (1) specifying parametric functional forms for the treatment effect over time, or (2) introducing categorical variables where each level represents the intervention effect for a specific exposure period. The non-parametric approach, which models time-varying treatment effects via categorical indicators, has been shown to reduce bias and achieve coverage probabilities close to the nominal level \citep{nickless18}. 
Such models provide greater flexibility for capturing time-varying effects, offering a more accurate representation of the intervention's impact over time.
However, as the number of exposure periods increases, the number of levels in the categorical variable, and consequently, the number of parameters to estimate also grows. This can lead to wider confidence intervals and a potential loss of efficiency due to increased variability in the estimates \citep{nickless18, maleyeff23}. To address these challenges, Maleyeff et al. \cite{maleyeff23} proposed a new model that borrows information across exposure periods by incorporating random effects to account for time-varying treatment effects, offering a balance between flexibility and parsimony \citep{maleyeff23}. By introducing a working variance component parameterization to capture treatment effect heterogeneity over time while reducing model complexity, this approach mitigates the trade-off between accuracy and efficiency in SW-CRTs. 

Despite the burgeoning literature on addressing time-varying treatment effects in stepped wedge designs, to the best of our knowledge, there has been no prior efforts in investigating the implications of time-varying treatment effects in stepped wedge designs with multiple interventions. In recent years, the implementation of multiple interventions within a single trial is increasingly common in many SW-CRTs, although such multiple-intervention trials are often analyzed under a constant treatment effect framework \citep{lyons17,insight,grayling19,zhang20,sundin22}. To fill in this methodological gap, we develop two models that go beyond constant treatment effects and are suitable for time-varying treatment effects in the context of stepped wedge trials with multiple interventions: (1) a non-parametric, fixed time-varying treatment effect model extending Hughes et al\cite{hgh15}, and (2) a working model with a random treatment effect over exposure time extending Maleyeff et al.\cite{maleyeff23} In the general setting of a stepped wedge trial with multiple interventions, we derive the formula for the expected value of the constant treatment effect estimator when treatment effects, in fact, vary across exposure time periods. When applied to the special case of a single-intervention stepped wedge trial, our results recover the findings established in \cite{kenny22} as a special case. Via simulation studies, we evaluate the performance of a constant effect model, a non-parametric, fixed time-varying treatment effect model, and a random-effects model under both constant and varying exposure-period-specific treatment effect patterns, 
in the context of concurrent and factorial stepped wedge cluster randomized trials, to illustrate the implications of ignoring exposure time treatment effect heterogeneity in this more complex setting. Additionally, we compare the statistical power of the concurrent versus factorial stepped wedge designs empirically through simulation studies. Finally, we apply the developed models to the Prognosticating Outcomes and Nudging Decisions in the Electronic health Record (PONDER) study, which was a factorial SW-CRT with multiple interventions.

The rest of this article is organized as follows. In Section~\ref{notation}, we introduce three common designs for stepped wedge cluster randomized trials with multiple interventions. Section~\ref{model} reviews the constant treatment effect model and proposes two time-varying treatment effect models. Section~\ref{results} presents the derivation of the expected value formulas for the constant effect estimator when treatment effects are heterogeneous over exposure periods. Section~\ref{sec:simu} reports the results of two simulation studies, and Section~\ref{sec: data analysis} applies the proposed models to the PONDER study. All technical derivations are provided in the Supplemental Materials.

\section{Variations of Stepped Wedge Trial design and the implied time-varying treatment effect structures} \label{notation}
In this section, we begin by discussing SW-CRT designs with a single intervention exhibiting time-varying treatment effects.  We then introduce three common variants involving multiple interventions: concurrent, supplementation, and factorial designs, following the classification introduced in Lyon et al.\cite{lyons17} Consider a trial with $T$ equally spaced study time periods, $S$ sequences, and $I$ clusters evenly allocated across sequences, with measurements taken at each time period. Each sequence is defined by the time period in which its assigned cluster group crosses over to the intervention and the specific intervention it receives. 
Without loss of generality, we will focus on one cluster per sequence to simplify the discussion. We use $i=1,\dots,I$ to index the clusters and $j=1,\dots,T$ to index the time periods. In designs with multiple interventions, we define $m$ as the total number of interventions and use $k=1,\ldots,m$ to index each intervention. The number of sequences is denoted by $S$, and in a standard single-intervention stepped wedge design, we assume $S=T-1$. When multiple interventions are considered, the total number of clusters would be determined by $I=m(T-1)$ in a concurrent design framework (introduced in Section \ref{sec: Concurrent design}). 

\subsection{Single-intervention stepped wedge design}

A standard SW-CRT with a single intervention enrolls $S=T-1$ sequences,  where each sequence starts in the control period, and in a randomly selected sequence, each cluster (or group of clusters) crosses over to the intervention at subsequent, and generally standard time periods, i.e., steps. Figure~\ref{fig:single} illustrates this design for $T=5$, with $I=4$ clusters evenly distributed across $S=4$ sequences.
Let $\bdelta=(\delta_1, \dots ,\delta_{T-1})'$ denote the treatment effects at exposure periods $1$ to $T-1$, with $\delta_0=0$ for the control period. 
The treatment effect received by cluster $i$ at study time $j$ follows
 $\delta_{\max\{j-i,0\}}$,
for $j=1,\ldots,T$. If $j\leq i$,  cluster $i$ remains in the control period, and thus the treatment effect is zero by definition. 
\begin{figure}[htbp!]
\setlength{\unitlength}{0.1in} 
\centering 
\begin{picture}(26,12) 
\setlength\fboxsep{0pt}

\put(1,8){\framebox(4,1.5){0}}
\put(7,8){\colorbox{gray!40}{\framebox(4,1.5){$\delta_1$}}}
\put(13,8){\colorbox{gray!40}{\framebox(4,1.5){$\delta_2$}}}
\put(19,8){\colorbox{gray!40}{\framebox(4,1.5){$\delta_3$}}} 
\put(25,8){\colorbox{gray!40}{\framebox(4,1.5){$\delta_{4}$}}}

\put(1,6){\framebox(4,1.5){0}}
\put(7,6){\framebox(4,1.5){0}}
\put(13,6){\colorbox{gray!40}{\framebox(4,1.5){$\delta_1$}}}
\put(19,6){\colorbox{gray!40}{\framebox(4,1.5){$\delta_2$}}}
\put(25,6){\colorbox{gray!40}{\framebox(4,1.5){$\delta_3$}}}

\put(1,4){\framebox(4,1.5){0}}
\put(7,4){\framebox(4,1.5){0}}
\put(13,4){\framebox(4,1.5){0}}
\put(19,4){\colorbox{gray!40}{\framebox(4,1.5){$\delta_1$}}}
\put(25,4){\colorbox{gray!40}{\framebox(4,1.5){$\delta_2$}}}

\put(1,2){\framebox(4,1.5){0}}
\put(7,2){\framebox(4,1.5){0}}
\put(13,2){\framebox(4,1.5){0}}
\put(19,2){\framebox(4,1.5){0}}
\put(25,2){\colorbox{gray!40}{\framebox(4,1.5){$\delta_1$}}}
\put(-3,2.5){$i=4$}
\put(-3,4.5){$i=3$}
\put(-3,6.5){$i=2$}
\put(-3,8.5){$i=1$}

\put(1.5,10.4){$j=1$}\put(7.5,10.4){$j=2$}
\put(13.5,10.4){$j=3$}\put(19.5,10.4){$j=4$}\put(25.5,10.4){$j=5$}
\end{picture}
  \caption{A schematic illustration of a stepped-wedge CRT with a single intervention,  $I=4$ clusters, and $T=5$ periods. Each white cell with a $0$ entry represents a cluster-period under the control condition. Each gray cell represents a cluster-period under the intervention condition, where $\delta_{j-i}$ denotes the treatment effect at the $(j-i)$-th exposure period for $i < j \leq T$. }
\label{fig:single}
\end{figure}

\subsection{Multiple-intervention stepped wedge design}
\subsubsection{Concurrent design}\label{sec: Concurrent design}
In a concurrent stepped wedge design, multiple interventions are introduced simultaneously across different clusters, and each cluster receives one of the interventions exclusively. In the context of time-varying treatment effects, we consider a design scheme with $m$ interventions and $I=m(T-1)$ clusters, where the first $T-1$ clusters are assigned the first intervention, the next $T-1$ clusters are assigned the second intervention, and so forth. More formally, cluster $i$ receives intervention $k$ if $(k-1)(T-1)+1\leq i\leq k(T-1)$. Each intervention is rolled out following the structure of a standard single-intervention stepped wedge design, and the study concludes once all clusters have received one intervention. Figure~\ref{fig:concurrent} illustrates a concurrent design with $m=2$ interventions and $T=5$ periods.

Write $\bdelta_{k}=(\delta_{k,1}, \dots, \delta_{k,T-1})'$ for the exposure-time-specific treatment effect vector for  intervention $k$. At exposure period 0, the treatment effect is zero for all interventions, i.e. $\delta_{k,0}=0$ for $1\leq k\leq m$. Under this design, the exposure time under intervention $k$ for the $i$-th cluster at study time $j$ is given by 
$$\max\{j-[i-(k-1)(T-1)],0\},$$ 
where $1\leq i-(k-1)(T-1)\leq T-1$.

\begin{figure}[ht]
\setlength{\unitlength}{0.1in} 
\centering 
\begin{picture}(26,20) 
\setlength\fboxsep{0pt}
\put(1,16){\framebox(4,1.5){0}}
\put(7,16){\colorbox{gray!40}{\framebox(4,1.5){$\delta_{1,1}$}}}
\put(13,16){\colorbox{gray!40}{\framebox(4,1.5){$\delta_{1,2}$}}}
\put(19,16){\colorbox{gray!40}{\framebox(4,1.5){$\delta_{1,3}$}}}
\put(25,16){\colorbox{gray!40}{\framebox(4,1.5){$\delta_{1,4}$}}}

\put(1,14){\framebox(4,1.5){0}}
\put(7,14){\framebox(4,1.5){0}}
\put(13,14){\colorbox{gray!40}{\framebox(4,1.5){$\delta_{1,1}$}}}
\put(19,14){\colorbox{gray!40}{\framebox(4,1.5){$\delta_{1,2}$}}}
\put(25,14){\colorbox{gray!40}{\framebox(4,1.5){$\delta_{1,3}$}}}

\put(1,12){\framebox(4,1.5){0}}
\put(7,12){\framebox(4,1.5){0}}
\put(13,12){\framebox(4,1.5){0}}
\put(19,12){\colorbox{gray!40}{\framebox(4,1.5){$\delta_{1,1}$}}}
\put(25,12){\colorbox{gray!40}{\framebox(4,1.5){$\delta_{1,2}$}}}

\put(1,10){\framebox(4,1.5){0}}
\put(7,10){\framebox(4,1.5){0}}
\put(13,10){\framebox(4,1.5){0}}
\put(19,10){\framebox(4,1.5){0}}
\put(25,10){\colorbox{gray!40}{\framebox(4,1.5){$\delta_{1,1}$}}}

\put(1,8){\framebox(4,1.5){0}}
\put(7,8){\colorbox{yellow!40}{\framebox(4,1.5){$\delta_{2,1}$}}}
\put(13,8){\colorbox{yellow!40}{\framebox(4,1.5){$\delta_{2,2}$}}}
\put(19,8){\colorbox{yellow!40}{\framebox(4,1.5){$\delta_{2,3}$}}}
\put(25,8){\colorbox{yellow!40}{\framebox(4,1.5){$\delta_{2,4}$}}}

\put(1,6){\framebox(4,1.5){0}}
\put(7,6){\framebox(4,1.5){0}}
\put(13,6){\colorbox{yellow!40}{\framebox(4,1.5){$\delta_{2,1}$}}}
\put(19,6){\colorbox{yellow!40}{\framebox(4,1.5){$\delta_{2,2}$}}}
\put(25,6){\colorbox{yellow!40}{\framebox(4,1.5){$\delta_{2,3}$}}}

\put(1,4){\framebox(4,1.5){0}}
\put(7,4){\framebox(4,1.5){0}}
\put(13,4){\framebox(4,1.5){0}}
\put(19,4){\colorbox{yellow!40}{\framebox(4,1.5){$\delta_{2,1}$}}}
\put(25,4){\colorbox{yellow!40}{\framebox(4,1.5){$\delta_{2,2}$}}}

\put(1,2){\framebox(4,1.5){0}}
\put(7,2){\framebox(4,1.5){0}}
\put(13,2){\framebox(4,1.5){0}}
\put(19,2){\framebox(4,1.5){0}}
\put(25,2){\colorbox{yellow!40}{\framebox(4,1.5){$\delta_{2,1}$}}}

\put(-3,16.5){$i=1$}\put(-3,14.5){$i=2$}\put(-3,12.5){$i=3$}\put(-3,10.5){$i=4$}
\put(-3,8.5){$i=5$}\put(-3,6.5){$i=6$}\put(-3,4.5){$i=7$}\put(-3,2.5){$i=8$}
\put(1.5,18.4){$j=1$}\put(7.5,18.4){$j=2$}\put(13.5,18.4){$j=3$}\put(19.5,18.4){$j=4$}\put(25.5,18.4){$j=5$}

\put(-4,13.5){\makebox(0,0)[c]{$\left\{\rule{0pt}{3\unitlength}\right.$}}
\put(-9,13.5){\makebox(0,0)[c]{Intervention 1}}

\put(-4,5.7){\makebox(0,0)[c]{$\left\{\rule{0pt}{3\unitlength}\right.$}}
\put(-9,6){\makebox(0,0)[c]{Intervention 2}}
\end{picture}
\caption{A schematic illustration of a concurrent design with $m=2$ interventions, $T=5$ periods, and $4$ distinct intervention sequences for each intervention. Clusters $1$–$4$ are assigned to intervention $1$, and clusters $5$–$8$ are assigned to intervention $2$. White cells with a $0$ entry represent cluster-periods under the control condition. Gray cells represent cluster-periods under intervention 1, and yellow cells under intervention 2.} 
\label{fig:concurrent} 
\end{figure}

\subsubsection{Supplementation design}
The supplementation design extends the standard single-intervention stepped wedge design by introducing a second intervention as an add-on to the first. This design facilitates the evaluation of both the separate effect of the first intervention and the combined effect of both interventions.  Under the assumption of time-invariant treatment effects and an additive combined treatment effect, where the joint effect of both interventions equals the sum of their individual effects, the marginal effect of the second intervention can also be identified. However, if treatment effects vary over time, the exposure-time-specific treatment effects become unidentifiable, even under the additive combined effect assumption. 

For illustration, consider a supplementation stepped wedge design depicted in Figure \ref{fig:supp}, with $m=2$ interventions, $I=3$ clusters and $T=5$ time periods, assuming an additive combined treatment effect. When treatment effects vary with exposure time, there are seven distinct treatment effects, including $\delta_{1,1}$ to $\delta_{1,4}$ for intervention $1$, and $\delta_{2,1}$ to $\delta_{2,3}$ for intervention $2$. In this case, one may identify $\delta_{1,1}$, $\delta_{1,2} + \delta_{2,1}$, $\delta_{1,3} + \delta_{2,2}$, and $\delta_{1,4} + \delta_{2,3}$, but not all exposure-time-specific effects without imposing parametric modeling assumptions on the effect pattern. If the treatment effect of intervention 1 is assumed to be constant over time, then $\delta_{1,1}=\delta_{1,2}$ and hence $\delta_{2,1}$ can be identified. 

\begin{figure}[htbp!]
\setlength{\unitlength}{0.1in} 
\centering 
\begin{picture}(36,8) 
\setlength\fboxsep{0pt}
\put(1,6){\framebox(6,1.5){0}}
\put(9,6){\colorbox{gray!40}{\framebox(6,1.5){$\delta_{1,1}$}}}
\put(17,6){\colorbox{orange!40}{\framebox(6,1.5){$\delta_{1,2}+\delta_{2,1}$}}}
\put(25,6){\colorbox{orange!40}{\framebox(6,1.5){$\delta_{1,3}+\delta_{2,2}$}}}
\put(33,6){\colorbox{orange!40}{\framebox(6,1.5){$\delta_{1,4}+\delta_{2,3}$}}}

\put(1,4){\framebox(6,1.5){0}}
\put(9,4){{\framebox(6,1.5){0}}}
\put(17,4){\colorbox{gray!40}{\framebox(6,1.5){$\delta_{1,1}$}}}
\put(25,4){\colorbox{orange!40}{\framebox(6,1.5){$\delta_{1,2}+\delta_{2,1}$}}}
\put(33,4){\colorbox{orange!40}
{\framebox(6,1.5){$\delta_{1,3}+\delta_{2,2}$}}}

\put(1,2){\framebox(6,1.5){0}}
\put(9,2){\framebox(6,1.5){0}}
\put(17,2){\framebox(6,1.5){0}}
\put(25,2){\colorbox{gray!40}{\framebox(6,1.5){$\delta_{1,1}$}}}
\put(33,2){\colorbox{orange!40}{\framebox(6,1.5){$\delta_{1,2}+\delta_{2,1}$}}}
\put(1,6){\framebox(6,1.5){0}}
\put(9,6){\colorbox{gray!40}{\framebox(6,1.5){$\delta_{1,1}$}}}
\put(17,6){\colorbox{orange!40}{\framebox(6,1.5){$\delta_{1,2}+\delta_{2,1}$}}}
\put(25,6){\colorbox{orange!40}{\framebox(6,1.5){$\delta_{1,3}+\delta_{2,2}$}}}
\put(33,6){\colorbox{orange!40}{\framebox(6,1.5){$\delta_{1,4}+\delta_{2,3}$}}}

\put(1,4){\framebox(6,1.5){0}}
\put(9,4){{\framebox(6,1.5){0}}}
\put(17,4){\colorbox{gray!40}{\framebox(6,1.5){$\delta_{1,1}$}}}
\put(25,4){\colorbox{orange!40}{\framebox(6,1.5){$\delta_{1,2}+\delta_{2,1}$}}}
\put(33,4){\colorbox{orange!40}
{\framebox(6,1.5){$\delta_{1,3}+\delta_{2,2}$}}}

\put(1,2){\framebox(6,1.5){0}}
\put(9,2){\framebox(6,1.5){0}}
\put(17,2){\framebox(6,1.5){0}}
\put(25,2){\colorbox{gray!40}{\framebox(6,1.5){$\delta_{1,1}$}}}
\put(33,2){\colorbox{orange!40}{\framebox(6,1.5){$\delta_{1,2}+\delta_{2,1}$}}}

\put(-3,2.5){$i=3$}\put(-3,4.5){$i=2$}\put(-3,6.5){$i=1$}
\put(2.5,8.4){$j=1$}
\put(10.5,8.4){$j=2$}
\put(18.5,8.4){$j=3$}
\put(26.5,8.4){$j=4$}
\put(34.5,8.4){$j=5$}

\end{picture}
  \caption{A schematic illustration of a supplementation  design with $m=2$ interventions, $I=3$ clusters and $T=5$ periods. Each white cell with a $0$ entry represents a cluster-period under the control condition, each gray cell receives intervention 1, and each orange cell receives both interventions, where $\delta_{1,j-i} + \delta_{2,j-i-1}$ represents an additive treatment effect at the $j$-th  period for $i+1 < j \leq T$. }
\label{fig:supp}
\end{figure}

\subsubsection{Factorial design}
A factorial stepped wedge design builds upon the concurrent and supplementation designs by allowing both interventions to be introduced independently and in combination across clusters and time periods. We consider a factorial design with $m$ interventions and $I = m(T-2)$ clusters. For instance, Figure \ref{fig:factorial design} illustrates a factorial design with $m=2$ and $T=5$, where each cluster transitions from a control state to one intervention, with the second intervention introduced at a later stage. Under the assumption of constant treatment effects, this design allows for the estimation of both the main effects of each intervention and their interaction effects. When treatment effects vary with exposure time, the exposure-time-specific main effects of each intervention remain identifiable, provided there are no interaction effects between them.

\begin{figure}[htbp!]
\setlength{\unitlength}{0.1in} 
\centering 
\begin{picture}(30,16) 
\setlength\fboxsep{0pt}
\put(1,12){\framebox(6,1.5){0}}
\put(9,12){\colorbox{gray!40}{\framebox(6,1.5){$\delta_{1,1}$}}}
\put(17,12){\colorbox{orange!40}{\framebox(6,1.5){$\delta_{1,2}+\delta_{2,1}$}}}
\put(25,12){\colorbox{orange!40}{\framebox(6,1.5){$\delta_{1,3}+\delta_{2,2}$}}}
\put(33,12){\colorbox{orange!40}{\framebox(6,1.5){$\delta_{1,4}+\delta_{2,3}$}}}

\put(1,10){\framebox(6,1.5){0}}
\put(9,10){\colorbox{yellow!40}{\framebox(6,1.5){$\delta_{2,1}$}}}
\put(17,10){\colorbox{orange!40}{\framebox(6,1.5){$\delta_{2,2}+\delta_{1,1}$}}}
\put(25,10){\colorbox{orange!40}{\framebox(6,1.5){$\delta_{2,3}+\delta_{1,2}$}}}
\put(33,10){\colorbox{orange!40}{\framebox(6,1.5){$\delta_{2,4}+\delta_{1,3}$}}}

\put(1,8){\framebox(6,1.5){0}}
\put(9,8){{\framebox(6,1.5){0}}}
\put(17,8){\colorbox{gray!40}{\framebox(6,1.5){$\delta_{1,1}$}}}
\put(25,8){\colorbox{orange!40}{\framebox(6,1.5){$\delta_{1,2}+\delta_{2,1}$}}}
\put(33,8){\colorbox{orange!40}
{\framebox(6,1.5){$\delta_{1,3}+\delta_{2,2}$}}}

\put(1,6){\framebox(6,1.5){0}}
\put(9,6){{\framebox(6,1.5){0}}}
\put(17,6){\colorbox{yellow!40}{\framebox(6,1.5){$\delta_{2,1}$}}}
\put(25,6){\colorbox{orange!40}{\framebox(6,1.5){$\delta_{2,2}+\delta_{1,1}$}}}
\put(33,6){\colorbox{orange!40}
{\framebox(6,1.5){$\delta_{2,3}+\delta_{1,2}$}}}

\put(1,4){\framebox(6,1.5){0}}
\put(9,4){\framebox(6,1.5){0}}
\put(17,4){\framebox(6,1.5){0}}
\put(25,4){\colorbox{gray!40}{\framebox(6,1.5){$\delta_{1,1}$}}}
\put(33,4){\colorbox{orange!40}{\framebox(6,1.5){$\delta_{1,2}+\delta_{2,1}$}}}

\put(1,2){\framebox(6,1.5){0}}
\put(9,2){\framebox(6,1.5){0}}
\put(17,2){\framebox(6,1.5){0}}
\put(25,2){\colorbox{yellow!40}{\framebox(6,1.5){$\delta_{2,1}$}}}
\put(33,2){\colorbox{orange!40}{\framebox(6,1.5){$\delta_{2,2}+\delta_{1,1}$}}}

\put(-3,12.5){$i=1$}
\put(-3,10.5){$i=2$}
\put(-3,8.5){$i=3$}
\put(-3,6.5){$i=4$}
\put(-3,4.5){$i=5$}
\put(-3,2.5){$i=6$}

\put(2.5,14.4){$j=1$}
\put(10.5,14.4){$j=2$}
\put(18.5,14.4){$j=3$}
\put(26.5,14.4){$j=4$}
\put(34.5,14.4){$j=5$}
\end{picture}
\caption{A factorial stepped wedge design with $m=2$ interventions, $I=6$ clusters and $T=5$ time periods. White cells represent control periods, gray cells receive intervention 1,  yellow cells receive intervention 2, and orange cells receive a combination of two interventions, where an additive treatment effect at time $j$ is represented by either $\delta_{1,j-i} + \delta_{2,j-i-1}$ or $\delta_{2,j-i} + \delta_{1,j-i-1}$, for $i+1 < j \leq T$.}
\label{fig:factorial design} 
\end{figure}

\section{Expanding Linear Mixed Models to Address Time-Varying Treatment Effects in Multiple-Intervention Stepped Wedge Designs}\label{model}
In this section, we describe three analytical models for stepped wedge designs with multiple interventions, including a constant treatment effect model \citep{insight,lyons17,sundin22}, a time-varying fixed treatment effect model\citep{overview, kenny22}, and a random treatment effect model \citep{maleyeff23}. 

To formalize these models, we introduce the following notations. Let $\bone$ and $\bzero$ represent column vectors of ones and zeros, respectively, with their transposed row vectors denoted by $\bone'$ and $\bzero'$. The matrix of all ones is given by $\bJ=\bone\bone'$, and a matrix of zeros by $\bO$.
Let $\bI$ denote the identity matrix. 
We assume each notation is of appropriate dimension in the relevant context. In cases of ambiguity, subscripts are used to specify dimensions.  For instance, $\bone_3$ indicates a column vector of three ones, $\bzero_4'$ is a row vector of four zeros, and $\bO_{2\times 3}$ is a $2\times 3$ matrix of zeros.  When referencing treatment effects with multiple indices, we separate them with commas for clarity; for instance, $\delta_{k,0}$ refers to the treatment effect of the $k$-th intervention at exposure time $0$. 
Vectors or matrices with zero dimensions are undefined and considered empty. 
Define $n_{ij}$ as the cluster-period size for cluster $i$ in time period  $j$.
Let $x_{kij}$ indicate whether the $k$-th intervention is applied in cluster $i$ at time $j$, with $x_{kij}=1$ denoting intervention and $x_{kij}=0$ denoting control. Since all clusters begin in the control condition, we have $x_{ki1}=0$ for all $i$ and $k$. Finally, let $y_{ijs}$ represent  the continuous outcome for individual $s$ in cluster $i$ at time period $j$. To focus ideas, we will also focus on the class of random-intercept models, and the extensions of these models to more complicated random-effects structures,\cite{overview} are generally straightforward. We will return to a discussion of these extensions in Section \ref{summary}.

\subsection{Constant treatment effect model}
We first consider the constant treatment effect model, which assumes a fixed treatment effect for each intervention. The individual-level outcome model is specified as
\begin{equation}\label{eq: constant model}
    y_{ijs}=\beta_j+x_{1ij} \theta_1 +\cdots + x_{mij} \theta_m+\alpha_i+\epsilon_{ijs},
\end{equation}
where $\beta_j$ represents the fixed effect for time period $j$, $\theta_k$ denotes the treatment effect of intervention $k$, and $\alpha_i 
{\sim} \mathcal{N}(0, \sigma^2_\alpha)$ is a cluster-specific random intercept.
The residual errors $\epsilon_{ijs} {\sim} \mathcal{N}(0, \sigma^2_\epsilon)$ are assumed to be independent across individuals and cluster-periods. No fixed intercept is included in model \eqref{eq: constant model} for identifiability, and $\alpha_i$ and $\epsilon_{ijs}$ are assumed to be mutually independent. The dependence among observations within the same cluster is quantified through a common intraclass correlation coefficient (ICC), defined as: $\sata/(\sata+\sate)$. 

Aggregating over $n_{ij}$ individuals in cluster $i$ at time $j$, the model for the cluster-period mean outcome is
\[
\bar{y}_{ij} \equiv \frac{1}{n_{ij}}\sum_{s=1}^{n_{ij}} y_{ijs}= \beta_j+x_{1ij} \theta_1 +\cdots + x_{mij} \theta_m+\alpha_i+\bar{\epsilon}_{ij}, \]
where $\bar{\epsilon}_{ij}=\frac{1}{n_{ij}}\sum_{s=1}^{n_{ij}}  \epsilon_{ijs} \sim \N(0,\sate/n_{ij})$. 
The variance of the cluster-period mean is therefore  $\Var (\bar{y}_{ij})=\sata+\sate/n_{ij}$. 

Define $\bar{\mathbf{y}}_i = (\bar{y}_{i1}, \ldots, \bar{y}_{iT})'$ as the vector of average outcomes for cluster $i$ across $T$ time periods, and let $\bar{\bme}_i=(\bar{\epsilon}_{i1}\ \cdots\ \bar{\epsilon}_{iT})'$ denote the corresponding vector of average residuals. The vector form of the model can then be expressed as:
\beq
\bar{\by}_{i} &=& \left( \begin{array}{c}
\beta_{1}\\
\cdots\\
\beta_{T} \end{array}\right)+\left( \begin{array}{c}
x_{1i1}\\
\cdots\\
x_{1iT} \end{array}\right) \theta_1+\ldots +\left( \begin{array}{c}
x_{mi1}\\
\cdots\\
x_{miT} \end{array}\right) \theta_m+\bone \alpha_i +\bar{\bme}_{i},
\enq
where $\bar{\bme}_i$ follows a multivariate normal distribution
with mean vector $\bzero$ and covariance matrix 
$\text{diag}\left( \frac{\sate}{n_{i1}}, \frac{\sate}{n_{i2}}, \dots, \frac{\sate}{n_{iT}} \right)$, where $\text{diag}(\cdot)$ denotes a diagonal matrix with the specified elements on the main diagonal and zeros elsewhere.
Let $\bth = (\theta_1, \ldots, \theta_m)'$ denote the vector of treatment effects, $\bbeta = (\beta_1, \ldots, \beta_T)'$ the vector of time effects, and $\mathbf{x}_{ki} = (x_{ki1}, \ldots, x_{kiT})'$ the vector of treatment allocations for intervention $k$ in cluster $i$. If intervention $k$ is introduced at time $j > 1$, then $x_{ki1} = \cdots = x_{ki, j-1} = 0$ and $x_{kij} = \cdots = x_{kiT} = 1$. Define the matrix $\mathbf{X}_i = (\mathbf{x}_{1i}, \ldots, \mathbf{x}_{mi})$.
The model can then be rewritten compactly as:
\begin{eqnarray}
\bar{\by}_{i} &=&\bbeta+\sum_{k=1}^m \bx_{ki}\theta_k+\bone \alpha_i+\bar{\bme}_i \nonumber \\ 
&=&\bbeta+\bX_i\bth+\bone \alpha_i+\bar{\bme}_i, \label{cmodel}
\end{eqnarray}
with the covariance structure given by 
$\Cov(\bar{\by}_{i})\equiv \bSigma_i=\sata\bJ+\text{diag}\left( \frac{\sate}{n_{i1}}, \frac{\sate}{n_{i2}}, \dots, \frac{\sate}{n_{iT}} \right)$.

\subsection{Time-varying fixed treatment effect model} \label{truemodel}

A more general linear mixed model allowing for exposure-time-specific treatment effects takes the form:
\begin{equation}\label{model: time-varying fixed}
   {y}_{ijs}= \beta_j+\sum_{k=1}^m x_{kij} \delta_{k, e_{kij}}+\alpha_i+{\epsilon}_{ijs},
\end{equation}
where $e_{kij}$ represents the cumulative exposure time under intervention $k$ for cluster $i$ at study time $j$. 
We note that $e_{kij}$ is a function of the corresponding $x_{kij}$ values, specifically given by $e_{kij} = \sum_{j'=1}^j x_{kij'}$.
In model \eqref{model: time-varying fixed},  $\delta_{k, e_{kij}}$ is the treatment effect as a function of the intervention level $k$ and exposure time $1\leq e_{kij} \leq T-1$, where $\delta_{k, 0} = 0$ for all $k$ by definition. When conducting statistical inference, researchers are often interested in the exposure-time-averaged treatment effect for each intervention $k$, defined as $\Delta_k= \sum_{j=1}^{T-1} \delta_{k, j}/(T-1)$. In model \eqref{eq: constant model}, the treatment effects are constant across each level of exposure time and hence $\Delta_k=\theta_k$.

\begin{figure}[h]
    \centering
    \includegraphics[width=1\linewidth]{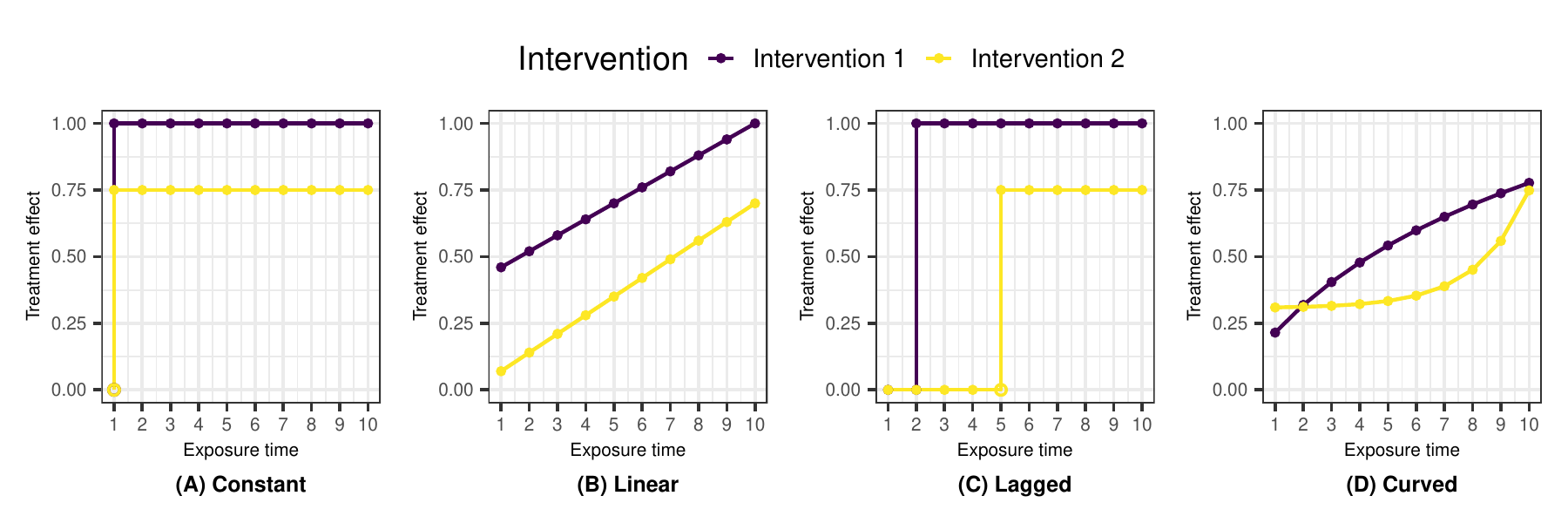}
    \caption{Treatment effect curves as a function of exposure time}
    \label{fig:effect-curves}
\end{figure}

In practical applications, treatment effects may exhibit diverse temporal patterns influenced by factors such as the nature of the intervention, population characteristics, and underlying mechanisms of change. Figure \ref{fig:effect-curves} provides examples of potential shapes of time-varying treatment effect curves that could arise in step-wedge designs with multiple interventions. For rapid-acting interventions, the treatment effect may reach its maximum immediately upon exposure and remain constant over time (Figure \ref{fig:effect-curves}A). In this scenario, model \eqref{model: time-varying fixed} reduces to model \eqref{eq: constant model} where the treatment effect $\delta_{k, e_{kij}}$ no longer depends on exposure time $e_{kij}$.  
For other interventions, treatment effects may grow proportionally with exposure duration, following a linear trajectory (Figure~\ref{fig:effect-curves}B). 
Some interventions, however, may not produce immediate effects after implementation; instead, a lag period may precede any measurable impact (Figure~\ref{fig:effect-curves}C).
More complex patterns include non-linear time-varying effects, such as logarithmic or exponential trajectories, are depicted in Figure \ref{fig:effect-curves}D. A logarithmic pattern shows a rapid initial increase in treatment effect, followed by a plateau, reflecting diminishing returns over time as saturation or constraints set in. In contrast, an exponential pattern begins with modest effects that accelerate, suggesting compounding benefits with continued exposure.

Let \(\bZ_{k,i}\) denote a \(T \times (T-1)\) indicator matrix for the exposure time under intervention \(k\) of cluster \(i\). For each row \(j = 1, \ldots, T\) and column \(e = 1, \ldots, T-1\), the \((j, e)\)-th element of \(\bZ_{k,i}\) is defined as:  
\[
[\bZ_{k,i}]_{j,e} = 
\begin{cases} 
1, & \text{if } e_{kij} = e, \\
0, & \text{otherwise}.
\end{cases}
\]
In the concurrent design described in Section~\ref{sec: Concurrent design}, when \((k-1)(T-1)+1 \leq i \leq k(T-1)\), the exposure time is given by:  
\[
e_{kij} = \max\{j + (k-1)(T-1) - i, 0\},
\]
and the matrix \(\bZ_{k,i}\)  takes the block structure:  
\[
\bZ_{k,i} = 
\begin{pmatrix*}[l]
\bO_{l \times (T-l)} & \bO_{l\times (l-1)}\\
 \bI_{T-l} & \bO_{(T-l) \times (l-1)} 
\end{pmatrix*},
\]
 where $l=i-(k-1)(T-1)$ and  $l=1,\ldots,T-1$. 
For clusters outside this intervention range, i.e., \(i \leq (k-1)(T-1)\) or \(i > k(T-1)\), we set \(\bZ_{k,i} = \bO\).
Let \(\bZ_i = (\bZ_{1,i}\ \cdots\ \bZ_{m,i})\) be the treatment effect indicator matrix for  cluster $i$ with dimensions \(T \times m(T-1)\). As defined in Section \ref{sec: Concurrent design}, let \(\bdelta_k = (\delta_{k,1}\ \cdots\ \delta_{k,T-1})'\) represent the vector of fixed treatment effects for intervention \(k\) across exposure periods, and  $\bdelta = (\bdelta_1'\ \cdots\ \bdelta_m')'$.
The time-varying fixed treatment effect model for the vector \(\bar{\by}_i\) is then expressed as:  
\begin{align}
\bar{\by}_i &= \bbeta + \sum_{k=1}^m \bZ_{k,i} \bdelta_k + \bone \alpha_i + \bar{\bme}_i \nonumber \\
            &= \bbeta + \bZ_i \bdelta + \bone \alpha_i + \bar{\bme}_i. \label{model: fixed time-varying vector form}
\end{align}
From this model, we have $\E(\bar{\by}_i) = \bbeta + \bZ_i \bdelta$ and the same covariance structure 
$\Cov(\bar{\by}_{i}) =\sata\bJ+\text{diag}\left( \frac{\sate}{n_{i1}}, \frac{\sate}{n_{i2}}, \dots, \frac{\sate}{n_{iT}} \right)$ as in the constant treatment effect model \eqref{eq: constant model}.
In this time-varying fixed effects framework, the model parameters consist of the fixed effects \(\{\bbeta, \bdelta\}\) and the variance components \(\{\sata, \sate\}\).

\subsection{Random treatment effect model}
While the time-varying fixed treatment effect model described in Section~\ref{truemodel} provides flexibility in modeling exposure-time-specific treatment effects, the number of parameters to estimate increases linearly with the number of exposure periods. This can lead to cost in degrees of freedom and precision, particularly in scenarios with limited sample sizes and a large number of periods.

An alternative approach to account for the time-varying treatment effects is to model these  effects as random effects\cite{maleyeff23}, through a parametric working random effect distribution assumption. 
Specifically, for each intervention $k$ exposed for $e_{kij}$ periods, we decompose its treatment effect  as $$\delta_{k,e_{kij}} = \mu_k + \gamma_{k,e_{kij}},$$ where $\mu_k$ represents the overall  mean effect of intervention $k$ across all exposure time periods, and $\gamma_{k,e_{kij}}$ denotes the random deviation from this mean effect at exposure time $e_{kij}$.
The corresponding random treatment effect
model is given by
\begin{equation}\label{model: time-varying random}
   {y}_{ijs}= \beta_j+\sum_{k=1}^m x_{kij} (\mu_k + \gamma_{k,e_{kij}})+\alpha_i+{\epsilon}_{ijs},
\end{equation}
Define the random vector for deviations as $\boldsymbol{\gamma}_{k} = (\gamma_{k,1},\dots,\gamma_{k,T-1})'$. The cluster-period means can then be expressed as
\begin{equation}\label{model: random}
\bar{\by}_{i} = \bbeta+\sum_{k=1}^m \bZ_{ki}(\bone\mu_k+\boldsymbol{\gamma}_{k})+\bone \alpha_i+\bar{\bme}_i, 
\end{equation}
where the random vectors $\boldsymbol{\gamma}_{k}$ are assumed to be mutually independent across all 
 $k$ and independent of
$\alpha_i$'s and ${\bme}_i$'s. 
Assuming $\boldsymbol{\gamma}_{k}$ follows a multivariate normal distribution
$\N(\boldsymbol{0},\sat_k\bI)$, the treatment effects of each intervention $k$ across various exposure
periods then share the same mean $\mu_k$. The parameters in model \eqref{model: random} include
$\{ \bbeta,\mu_k, \sat_k,\sata,\sate\}$. It is important to note that model~\eqref{model: random} is a working model, in the sense that it relies on the assumption of exchangeability in the random deviations $\gamma_{k,e_{kij}}$, with a common variance structure across exposure periods. While this assumption simplifies estimation and can be attractive especially for estimating the exposure time-averaged treatment effect,\cite{maleyeff23} it may not fully capture complex temporal dependencies or heterogeneity in treatment effects across different exposure periods.

\section{Large-Sample Behavior of the constant treatment effect model under time-varying fixed treatment effects}\label{results}

In this section, we investigate the properties of the treatment effect estimators derived from the constant treatment effect model \eqref{cmodel} when applied to different stepped wedge designs with multiple interventions. 
While these estimators are unbiased under constant treatment effects, their behavior under time-varying treatment effects requires careful examination. 
Without loss of generality, we assume a constant cluster-period size, $n_{ij} = n$, for all $i$ and $j$. 
We first derive general expressions for potential bias when the true underlying effects vary with time according to model \eqref{model: fixed time-varying vector form}.  We then specialize these results to concurrent  and factorial stepped wedge trials, providing explicit formulas for bias quantification.

Consider a stepped wedge design where the data-generating model follows the time-varying treatment effect structure in \eqref{model: fixed time-varying vector form}. 
The generalized least squares estimators for the time effect vector $\bbeta$ and the treatment effect vector $\bth$ from fitting model \eqref{cmodel} are given by:
\beq
\left( \begin{array}{c}
\hat{\bbeta}\\
\hat{\bth}
\end{array}\right)&=&\left[ \sum_{i=1}^{I}  \left( \begin{array}{c}
\bI \\
\bX_i' \end{array}\right) {\bSigma_i}^{-1} \left(\bI \  \bX_i  \right)  \right]^{-1} \sum_{i=1}^{I}   \left( \begin{array}{c}
\bI \\
\bX_i' \end{array}\right)  {\bSigma_i}^{-1}\bar{\by}_{i}
\enq
Applying the inverse of a partitioned matrix, the treatment effect estimator $\hat{\bth}$ can be expressed as:
\begin{align*}
\hat{\bth}=
\left[ \sum_{i=1}^{I} \bX_i'\bSigma^{-1}_i\bX_i - \sum_{i=1}^{I}\bX_i'\bSigma_i^{-1}   \left( \sum_{i=1}^{I}\bSigma_{i}^{-1}\right)^{-1} \sum_{i=1}^{I}\bSigma^{-1}_i \bX_i \right]^{-1}
\left[  \sum_{i=1}^{I} \bX_i'  {\bSigma_i}^{-1}\bar{\by}_{i}-\sum_{i=1}^{I}\bX_i'\bSigma_i^{-1}\left( \sum_{i=1}^{I}\bSigma_{i}^{-1}\right)^{-1} \sum_{i=1}^{I}  {\bSigma_i}^{-1}\bar{\by}_{i}\right].     
\end{align*}
Define $\bar{\bX}=  \sum_{i=1}^{I} \bX_i/I$ and $\bar{\by}=\sum_{i=1}^{I}  \bar{\by}_{i}/I$. Then the estimator  $\h{\bth}$ simplifies to:
\bea
\hat{\bth}=\left[ \sum_{i=1}^{I} \bW(\bX_i) - \bS  \bar{\bX}\right]^{-1}\left[  \sum_{i=1}^{I} \bW (\bX_i,\bar{\by}_i) -\bS\bar{\by}\right], 
\label{gls}
\ena
where $\bW(\bX_i)=\bX_i'\bSigma_i^{-1}\bX_i$,  $ \bW(\bX_i,\bar{\by}_i)=\bX_i'{\bSigma_i}^{-1}\bar{\by}_{i}$,
and $\bS=\sum_{i=1}^{I}\bX_i'\bSigma_i^{-1}$.
\begin{theorem}\label{prop 1}
 Let $\bar{\bZ}=  \sum_{i=1}^{I} \bZ_i/I$ and $\bW(\bX_i,\bZ_i)=\bX_i'{\bSigma_i}^{-1}\bZ_{i}$. 
 Define the matrix $\bH=\left[ \sum_{i=1}^{I} \bW(\bX_i) - \bS  \bar{\bX} \right]^{-1} \left[  \sum_{i=1}^{I} \bW(\bX_i, \bZ_{i})-\bS\bar{\bZ}\right]$.
 If the true model follows the exposure-time dependent treatment effect model (\ref{model: fixed time-varying vector form}), then 
the expected value of the estimator $\h{\bth}$  in (\ref{gls}) satisfies
$\E (\h{\bth})=\bH\bdelta$. 
\end{theorem}In the following, we derive specific results for concurrent and factorial stepped wedge trials as applications of Theorem \ref{prop 1} to obtain more insights about the consequence of misspecifying the treatment effect structure in multiple-intervention stepped wedge designs. 

\subsection{Consequence of ignoring time-varying treatment effect in concurrent designs} 
In concurrent stepped wedge trials, each sequence receives one treatment exclusively.
Define $c=T(T-1)\left(3+b -2bT\right)/6$,  
$d=T\left[4T-2-3bT(T-1) \right]/(12m)$, and let $g= c-md=T(T-2)(2+b-bT)/12$, where $b = \sigma_{\alpha}^2/(T\sigma_{\alpha}^2+\sigma_{\epsilon}^2/n)$ is a design-dependent parameter related to the unknown variance components  $\sigma_{\alpha}^2$ and $\sigma_{\epsilon}^2$.
Let $\br$ be a vector whose $j$-th element  is 
$(T-j)\left[1+b(1-T-j)/2\right]$, and $\bv$ be a vector whose $j$-th element is
$(T-j)\left[ 1-bT+j/(T-1)\right]/(2m)$, for $j=1,\ldots,T-1$.
\begin{proposition}\label{thm}
For a concurrent stepped wedge trial with $m$ interventions, the weight matrix $\bH$ in Theorem~\ref{prop 1} takes the form:
\[\bH=\left[\frac{1}{c}\left(\bI_m+\frac{d}{g}\bJ_m\right)\right]\otimes \br'-\frac{1}{g}\bJ_m\otimes \bv'.\]
where $\otimes$ denotes the Kronecker product.  Consequently, the expected value of the constant effect estimator $\h{\theta}_k$ under intervention $k$ can be written as
\begin{equation}\label{Expectation of single treatment effect}
\E(\h{\theta}_k)=\frac{1}{c}\br'\bdelta_k+\frac{1}{g}\left(\frac{d}{c}\br'-\bv'\right)\bdelta_{*},
\end{equation}
for $k=1,\dots,m$, where $\bdelta_{*}=\sum_{k=1}^m \bdelta_k$ represents the aggregate of the treatment effects across all interventions at each exposure period. 
\end{proposition}
Expression \eqref{Expectation of single treatment effect} shows that the expected estimate $\E(\h{\theta}_k)$ is a weighted average of the associated
time-varying treatment effects $\bdelta_k$, along with an additional term independent of the specific intervention $k$. 
By decomposing $\bdelta_{*}$ into $\bdelta_k+\sum_{k'\neq k}^m \bdelta_{k'}$, we further obtain:
\[\E(\h{\theta}_k)=\left[\frac{1}{c}\left( 1+\frac{d}{g}\right)\br'-\frac{1}{g}\bv'\right]\bdelta_k+\frac{d}{cg}\br'\sum_{k'\neq k}\bdelta_{k'} -
\frac{1}{g}\bv'\sum_{k'\neq k}\bdelta_{k'}.\]
In general, the weight vector in front of $\bdelta_k$ is non-uniform and the additional term (which aggregates the contributions from interventions $k'\neq k$) does not vanish. Thus, unless additional constraints are imposed, $\hat{\theta}_k$ is generally a biased estimator for the true exposure-time-averaged treatment effect for the $k$th intervention,
\begin{equation}\label{causal estimand}
    \Delta_k= \sum_{j=1}^{T-1} \delta_{k, j}/(T-1).
\end{equation}
For the special case of a single intervention where $m=1$,  the expectation of the treatment effect reduces to
$$\E(\h{\theta})=6\sum_{j=1}^{T-1} \frac{w_j\delta_j}{T(T-1)(T-2)(2+b-bT)},$$ 
where the weights are given by $w_j=(T-j)\left[(b-1-b\,T)j +(1+b)(T -1)\right]$.
Note that the weights $w_j$ depend on the parameter $b$, which in turn is a function of the variance components $\sigma_{\alpha}^2$ and $\sigma_{\epsilon}^2$.

\begin{figure}[h]
    \centering
    \includegraphics[width=1\linewidth]{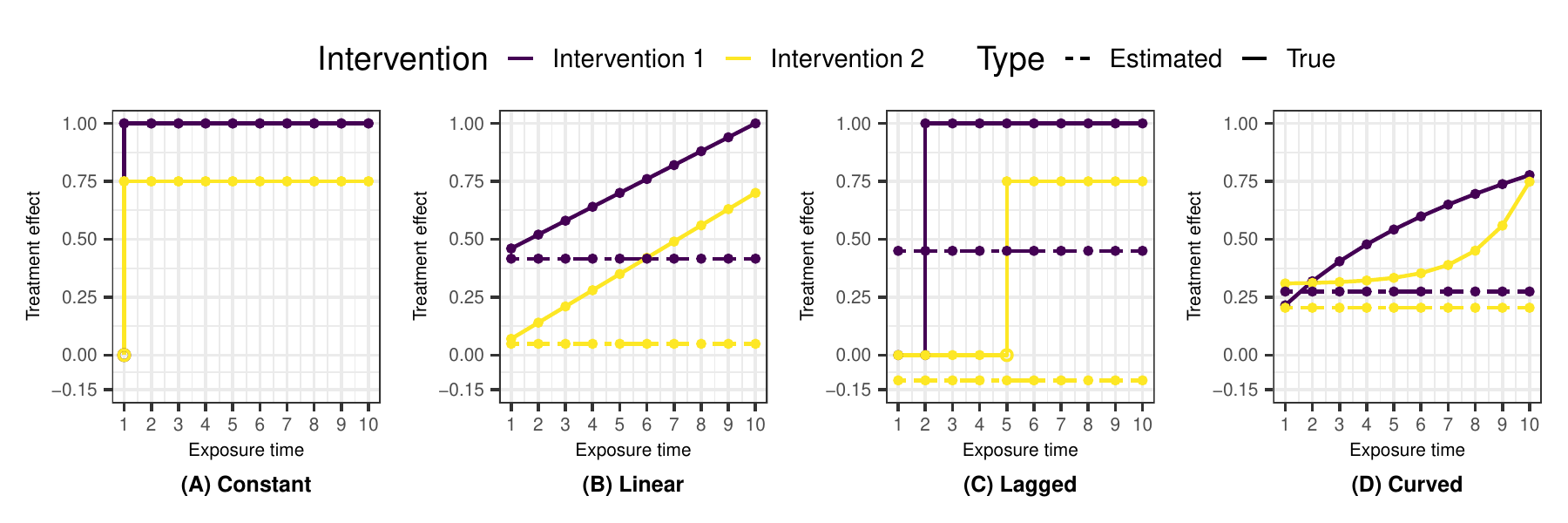}
\caption{Four possible true effect curves versus expected effect curves from a constant treatment effect model, for a concurrent design with $T = 11$ periods and $b = 1/T$. In Panel A, the dashed lines overlap with the solid lines.}
    \label{fig:bias-effect-curves concurrent}
\end{figure}

Figure \ref{fig:bias-effect-curves concurrent} displays the expected effect curve estimated from model \eqref{cmodel}  against the true effect curves under four different time-varying effect scenarios shown in Figure \ref{fig:effect-curves}.  As anticipated, when the true treatment effect is time-invariant, the constant effect estimator yields unbiased estimates of the exposure-time-averaged treatment effect, as shown in Panel A. However, when the true effects vary over time, imposing a constant effect assumption can lead to substantial bias. In some cases, this bias may even reverse the perceived direction of the effect. For instance, in Panel C, the estimated effect is negative despite the true effects being non-negative at all time points.

Finally, we identify conditions under which the constant effect estimator remains unbiased.
\bec\label{unbiased}
Suppose that in a concurrent stepped wedge trial with $m$ interventions, the treatment effect for a particular intervention $k$ is constant over time, i.e., 
$\bdelta_k=\bone\delta_k$. Then, the expected constant effect estimate for intervention $k$ is
\[\E(\h{\theta}_k)=\delta_k+\frac{1}{g}\left(\frac{d}{c}\br'-\bv'\right)\sum_{k^*\neq k}\bdelta_{k^*}.\]
That is, $\h{\theta}_k$ may still be biased to the exposure-time-averaged treatment effect for the $k$th intervention, when only the $k$th intervention exhibits a constant treatment effect pattern over exposure time (but other intervention effects are allowed to depend on exposure time). In particular, if \emph{all} interventions have constant treatment effects, so that $\bdelta_{k^*}=\bone\delta_{k^*}$ for all $1\leq k^*\leq m$, then we have $\E(\hat{\theta}_k)=\delta_k$, and the constant treatment effect estimator is unbiased for the exposure-time-averaged treatment effect, which remains a constant. 
\enc
 
\subsection{Consequence of ignoring time-varying treatment effect in factorial designs}
In a factorial design with two interventions, there are two time-varying treatment effect vectors, $\bdelta_1$ and $\bdelta_2$, each of length $T-1$. Hence, the dimension of the matrix $\bH$ in Theorem~\ref{prop 1} is $2\times 2(T-1)$.
Each cluster in the trial receives up to two interventions, and the 
 second intervention can start at a certain exposure period after the introduction of the first intervention.  
 That is, given $T$ time periods, the second intervention can begin as early as the second exposure period, or as late as the final exposure period, in a cluster that has already received the first intervention.
For instance, when $T=5$, the second intervention in the first sequence can start in period  $j=3$, as illustrated in Figure \ref{fig:factorial design}, or alternatively in periods $j=4$ or $j=5$. 

 \begin{proposition}\label{fthm}
 In a factorial stepped wedge trial, the weight matrix 
$\bH$  in Theorem~\ref{prop 1} exhibits the block structure $\bH=\left( \begin{matrix}
\bh_1' & \bh_2' \\
\bh_2'&\bh_1'
\end{matrix} \right)$, where each $\bh_k'$ is a row vector of length $T-1$, for $k=1,2$. 
 \end{proposition}
Following the above Proposition, the expected value of the constant treatment effect estimator is given by
 $$\E (\h{\bth})\equiv \left(\begin{matrix}
\E (\h{\theta}_1) \\
\E (\h{\theta}_2)
\end{matrix} \right)=\left( \begin{matrix}
\bh_1' & \bh_2' \\
\bh_2'&\bh_1'
\end{matrix} \right)\left( \begin{matrix}
 \bdelta_1 \\
\bdelta_2
\end{matrix} \right)=\left( \begin{matrix}
 \bh_1'\bdelta_1 +\bh_2'\bdelta_2\\
 \bh_2'\bdelta_1 +\bh_1'\bdelta_2
\end{matrix} \right).$$ 
The closed-form expression for $\bh_k$ depends on the design specifics.
Below we consider the simple case in Figure~\ref{t3} with \(T=3\) study periods. In this scenario, the design matrices for the four sequences are as follows:
 \[
\bX_1 = \begin{pmatrix}
0 & 0\\[0.5mm]
1 & 0\\[0.5mm]
1 & 1
\end{pmatrix},\quad
\bX_2 = \begin{pmatrix}
0 & 0\\[0.5mm]
0 & 0\\[0.5mm]
1 & 0
\end{pmatrix},\quad
\bX_3 = \begin{pmatrix}
0 & 0\\[0.5mm]
0 & 0\\[0.5mm]
0 & 1
\end{pmatrix},\quad
\bX_4 = \begin{pmatrix}
0 & 0\\[0.5mm]
0 & 1\\[0.5mm]
1 & 1
\end{pmatrix},
\]
and the corresponding treatment effect indicator matrices for these sequences are:
\[
\bZ_{11} = \bZ_{24} = \begin{pmatrix}
0 & 0\\[0.5mm]
1 & 0\\[0.5mm]
0 & 1
\end{pmatrix},\quad
\bZ_{12} = \bZ_{23} = \begin{pmatrix}
0 & 0\\[0.5mm]
0 & 0\\[0.5mm]
1 & 0
\end{pmatrix},
\bZ_{13} = \bZ_{22} = \mathbf{0}_{2\times 3},\quad
\bZ_{14} = \bZ_{21} = \begin{pmatrix}
0 & 0\\[0.5mm]
0 & 0\\[0.5mm]
1 & 0
\end{pmatrix}.
\]
Then the weight vectors in Proposition~\ref{fthm} simplify to
\begin{align*}
\bh_1'= \frac{1}{4(b-1)}\left(2b-3\  2b-1 \right),~~~~\bh_2'= \frac{1}{4(b-1)}\left( 1\ -1\right).
\end{align*}
Consequently, the expectation of the constant treatment effect estimator is given by
$$\left(\begin{matrix}
\E (\h{\theta}_1) \\
\E (\h{\theta}_2)
\end{matrix} \right)=\frac{1}{4(b-1)}\left(\begin{matrix*}[c]
2b-3& 2b-1 & 1& -1\\
1 & -1 &2b-3 & 2b-1
\end{matrix*} \right)\bdelta,$$ 
where $\bdelta=(\delta_{11} \ \delta_{12}\ \delta_{21}\ \delta_{22})'$ collects the time-varying effects for the two interventions across the two exposure periods.
For a large cluster size $n$, we have approximately $b=1/3$, and hence the weight matrix becomes $\bH= \left(\begin{matrix*}[c]
{7}/{8}& {1}/{8} & -{3}/{8}& {3}/{8}\\
-{3}/{8}& {3}/{8} &{7}/{8}& {1}/{8}
\end{matrix*} \right)$.  When $\bdelta = (1,-1,2,3)$, the true average treatment effects for the two interventions are $\Delta_1 = 0$ and $\Delta_1 =5/2$, while the constant treatment effect estimator yields $\E (\h{\theta}_1) = {9}/{8}$ and $\E (\h{\theta}_2) = {11}/{8}$, indicating that  the effect of the first intervention is overestimated, while the second intervention effect is underestimated.
\begin{figure}[htbp!]
\setlength{\unitlength}{0.1in} 
\centering 
\begin{picture}(20,10) 
\setlength\fboxsep{0pt}

\put(1,8){\framebox(6,1.5){0}}
\put(9,8){\colorbox{gray!40}{\framebox(6,1.5){$\delta_{1,1}$}}}
\put(17,8){\colorbox{orange!40}{\framebox(6,1.5){$\delta_{1,2}+\delta_{2,1}$}}}

\put(1,6){\framebox(6,1.5){0}}
\put(9,6){{\framebox(6,1.5){0}}}
\put(17,6){\colorbox{gray!40}{\framebox(6,1.5){$\delta_{1,1}$}}}

\put(1,4){\framebox(6,1.5){0}}
\put(9,4){\framebox(6,1.5){0}}
\put(17,4){\colorbox{yellow!40}{\framebox(6,1.5){$\delta_{2,1}$}}}

\put(1,2){\framebox(6,1.5){0}}
\put(9,2){\colorbox{yellow!40}{\framebox(6,1.5){$\delta_{2,1}$}}}
\put(17,2){\colorbox{orange!40}{\framebox(6,1.5){$\delta_{2,2}+\delta_{1,1}$}}}

\put(-3,8.5){$i=1$}
\put(-3,6.5){$i=2$}
\put(-3,4.5){$i=3$}
\put(-3,2.5){$i=4$}

\put(2.5,10.4){$j=1$}
\put(10.5,10.4){$j=2$}
\put(18.5,10.4){$j=3$}
\end{picture}
\caption{A factorial stepped wedge trial with $T=3$ time periods. In clusters 1 and 4, the second intervention starts at the second exposure period of the first intervention. \label{t3}}
\end{figure}

\begin{figure}[h]
    \centering
    \includegraphics[width=1\linewidth]{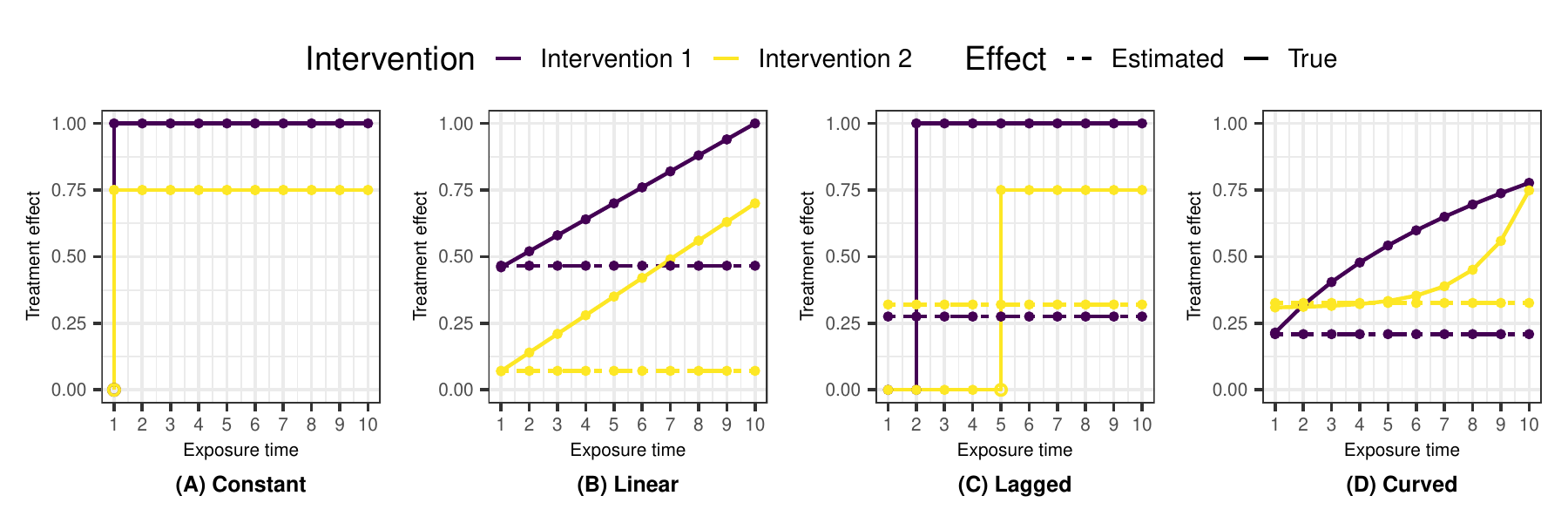}
\caption{Four possible true effect curves versus expected effect curves from a constant treatment effect model, for a factorial design with $T = 11$ periods and $b = 1/T$. In Panel A, the dashed lines overlap with the solid lines.}
    \label{fig:bias-effect-curves factorial}
\end{figure}

Figure \ref{fig:bias-effect-curves factorial} further illustrates the asymptotic bias under a factorial design with $T=11$ when the true effect curves follow the patterns shown in Figure \ref{fig:effect-curves}. The results are consistent with those observed in the concurrent design for the constant and linear effect patterns. However, for the lagged and curved effect scenarios, the magnitude of bias differs notably from that seen in the concurrent design. 

\section{Simulation Studies}\label{sec:simu}
This section presents two simulation studies under the ADEMP framework (Aims, Data-generating mechanisms, Estimands, Methods, Performance measures) as proposed by Morris et al. (2019) \cite{morris2019using}.

\subsection{Comparing different estimators}\label{subsec: simu compare estimators}

\subsubsection{Aims}
The goal of our first simulation study was to compare the performance of three models for estimating exposure-time-averaged treatment effects in stepped wedge trials. Specifically, we aimed to evaluate:
\textsf{Model A}, the constant treatment effect model; \textsf{Model B}, the
time-varying fixed treatment effect model; and \textsf{Model C}, the time-varying random treatment effect model. 
The vector form of each model is given in (\ref{cmodel}), (\ref{model: fixed time-varying vector form}) and (\ref{model: random}), respectively.

\subsubsection{Data-generating mechanisms}\label{Sim 1: DGP}

We conducted simulations for both concurrent trial and factorial trial designs with two interventions. The number of time periods was set to $T=5$ and $T=11$, with $2(T-1)$ clusters and a maximum exposure time of $T-1$ periods per intervention, where one cluster crossed over at each time point. Each cluster consisted of $n=30$ individuals, which mimics the average cluster-period size observed in our case study.
In the factorial trial design, the second intervention was introduced at the second exposure period of the first intervention when $T=5$ (as illustrated in Figure \ref{fig:factorial design}), and at the fourth exposure period of the first intervention when $T=11$. 

For each trial setting, data were generated from a linear mixed-effects model:
\[y_{ijs}=\beta_j+x_{1ij} g_1(e_{1ij})  + x_{2ij}  g_2(e_{2ij})+\alpha_i+\epsilon_{ijs},\]
where $ \alpha_i\sim \N(0,0.15)$ and $ \epsilon_{ijs}\sim \N(0,2.85)$. 
For each $T$, the fixed time effects $\bbeta$ 
  were chosen as $T$ equally spaced points from $0.1$ to $0.5$.
For each intervention $k=1,2$, the treatment effect $g_k(\cdot)$ was designed to represent four distinct exposure-time-specific patterns approximating the effect curves A to D in Figure \ref{fig:effect-curves}. 
All scenarios were calibrated to yield the same average treatment effect across exposure time.
We considered two average effect size scenarios: (1) small treatment effect: $\Delta_1=0.10$ for Intervention 1 and $\Delta_2=0.14$ for Intervention 2 when $T=5$, and $\Delta_1=0.10$ and $\Delta_2=0.13$ when $T=11$; (2) large treatment effect: $\Delta_1=0.28$ for Intervetion 1 and $\Delta_2=0.40$ for Intervetion 2 when $T=5$, and $\Delta_1=0.29$ and $\Delta_2=0.40$ when $T=11$. 
The specifications of the effect patterns are outlined below:
\begin{enumerate}
    \item \noindent In \textbf{Outcome Model A}, the treatment effect remains constant over exposure time:
$$g_k(e_{kij}) = \Delta_k.$$
\item In \textbf{Outcome Model B1}, the treatment effect increases linearly over exposure time. For each $T$,  elements of the time-varying treatment effect vector $\bdelta$ are $2(T-1)$ equally spaced points from $l$ to $u$:
\begin{align*}
    & g_k(e_{kij})=l+\frac{(u-l)}{2(T-1)-1} (e_{kij}+(k-1)(T-1)-1),
\end{align*}
where $l=0.08$ and $u=0.15$ for the small effect size scenario, and $l=0.24$ and $u=0.45$ for the large effect size scenario.

\item In \textbf{Outcome Model B2}, the treatment effect is initially delayed during the first $(T-1)/2$ exposure periods, becoming fully effective from exposure period $(T-1)/2+1$ onward:
 \begin{align*}
    & g_k(e_{kij})= 2\Delta_k\mathbbm{1}\{e_{kij} > \frac{T-1}{2}\}.
\end{align*}

\item In \textbf{Outcome Model B3}, the treatment effect is initially delayed during the first exposure period, becoming fully effective from the second exposure period onward:
 \begin{align*}
    & g_k(e_{kij})= \frac{T-1}{T-2}\Delta_k\mathbbm{1}\{e_{kij} > {1}\}.
\end{align*}

\item In \textbf{Outcome Model B4}, the treatment effect increases non-linearly over exposure time:
\begin{equation*}
 g_k(e_{kij}) = \Delta_k + f_k(e_{kij}) - \frac{1}{T-1} \sum_{t=1}^{T-1} f_k(e_{kij}), 
 \end{equation*}
 where
   $$  f_1(e_{1ij})= \Delta_1 \log\left\{\frac{T-1}{2}\left(1 + \frac{3(e_{1ij}-1)}{T-2}\right)\right\} \quad \text{and} \quad
     f_2(e_{2ij})= \Delta_2 \exp\left\{ -\frac{T-1}{2} + \frac{0.1 + \frac{T-1}{2}}{T-2} \cdot (e_{2ij}-1) \right\}.
$$
\end{enumerate}

 \subsubsection{Estimands}
The estimands of interest are the exposure-time-averaged treatment effects for each intervention, as defined in equation~\eqref{causal estimand}.

\subsubsection{Methods}

For each data generating process described above, 500 independent simulation replicates were generated. Within each replicate, 500 bootstrap samples were drawn by resampling individual data within each cluster-period.
For each simulated dataset, \textsf{Models A} and \textsf{B} were fit using the \textsf{lme} function from the \textsf{nlme} package, while \textsf{Model C} was fit using the \textsf{lmer} function from the \textsf{lme4} package of the statistical software \textsf{R}.
Confidence intervals for the exposure-time-averaged treatment effects were constructed using the bootstrap percentile method \citep{maleyeff23}.

\subsubsection{Performance measures}\label{sim1: results}

 We compared the estimated exposure-time-averaged treatment effect in terms of the empirical bias, the empirical standard error, the average of standard error estimates, 
the proportion of the coverage of the $95\%$ confidence intervals, and the average length of the confidence intervals. 

\begin{table}[ht]
\centering
\caption{Comparison of model estimates for the exposure-time-averaged treatment effects under different types of treatment effect heterogeneity across exposure time in a concurrent design, with a small effect size and 
 $T=5$ or $T=11$. We use $\hat{\Delta}$ to denote the estimate of the exposure-time-averaged treatment effect, $\hat{\E}(\cdot)$ and $\hat{\text{SD}}(\cdot)$ to denote the sample mean and sample standard deviation across $500$ simulations, and $\hat{\text{SE}}(\cdot)$ to denote the estimated standard error.} 
 \label{table: concurrent_small_effect}
\resizebox{\textwidth}{!}{
\begin{tabular}{cccccccccccc}
  \hline
    & &\multicolumn{5}{c}{$\Delta_1$} & \multicolumn{5}{c}{$\Delta_2$} \\ \cline{3-7} \cline{8-12}
    Outcome & Fitting &    &  & CI  &  &  &   & &CI  &  \\ 
 Model& Model & Bias &  $\hat{\text{SD}}(\hat{\Delta})$ &  Coverage (\%) & Length & $\hat{\E}(\hat{\text{SE}}(\hat{\Delta}))$  & Bias  & $\hat{\text{SD}}(\hat{\Delta})$ & Coverage (\%) & Length & $\hat{\E}(\hat{\text{SE}}(\hat{\Delta}))$ \\ 
  \hline
\multicolumn{12}{c}{\textbf{T=5, $\Delta_1=0.10$, $\Delta_2 = 0.14$}}\\
  A & A & -0.011 & 0.17 & 94.6 & 0.66 & 0.17 & -0.006 & 0.17 & 93.0 & 0.66 & 0.17 \\ 
  A & B & -0.020 & 0.20 & 94.0 & 0.79 & 0.20 & -0.011 & 0.20 & 95.4 & 0.80 & 0.20 \\ 
  A & C & -0.010 & 0.17 & 95.8 & 0.69 & 0.18 & -0.005 & 0.18 & 94.0 & 0.69 & 0.18 \\ 
  B1 & A & -0.009 & 0.16 & 95.0 & 0.66 & 0.17 & -0.009 & 0.16 & 96.0 & 0.66 & 0.17 \\ 
  B1 & B & 0.001 & 0.19 & 93.2 & 0.79 & 0.20 & 0.002 & 0.19 & 94.2 & 0.78 & 0.20 \\ 
  B1 & C & -0.011 & 0.16 & 95.6 & 0.69 & 0.18 & -0.010 & 0.16 & 96.0 & 0.69 & 0.18 \\ 
  B2 & A & -0.104 & 0.16 & 88.4 & 0.67 & 0.17 & -0.124 & 0.17 & 85.2 & 0.67 & 0.17 \\ 
  B2 & B & -0.001 & 0.19 & 95.6 & 0.80 & 0.20 & 0.003 & 0.20 & 93.6 & 0.80 & 0.20 \\ 
  B2 & C & -0.104 & 0.17 & 91.4 & 0.70 & 0.18 & -0.123 & 0.17 & 89.2 & 0.70 & 0.18 \\ 
  B3 & A & -0.059 & 0.17 & 92.2 & 0.66 & 0.17 & -0.070 & 0.16 & 92.2 & 0.66 & 0.17 \\ 
  B3 & B & -0.002 & 0.20 & 94.4 & 0.80 & 0.20 & -0.003 & 0.19 & 95.2 & 0.79 & 0.20 \\ 
  B3 & C & -0.060 & 0.17 & 93.4 & 0.69 & 0.18 & -0.071 & 0.16 & 93.8 & 0.69 & 0.18 \\ 
  B4 & A & -0.055 & 0.17 & 92.8 & 0.66 & 0.17 & -0.054 & 0.17 & 92.4 & 0.66 & 0.17 \\ 
  B4 & B & -0.007 & 0.19 & 94.6 & 0.79 & 0.20 & -0.006 & 0.19 & 96.2 & 0.79 & 0.20 \\ 
  B4 & C & -0.055 & 0.17 & 93.8 & 0.69 & 0.18 & -0.054 & 0.17 & 93.6 & 0.69 & 0.18 \\ 
   \hline
   \multicolumn{12}{c}{\textbf{T=11, $\Delta_1=0.10$, $\Delta_2 = 0.13$}}\\
 A & A & 0.003 & 0.07 & 94.6 & 0.26 & 0.07 & -0.001 & 0.07 & 94.6 & 0.26 & 0.07 \\ 
  A & B & 0.000 & 0.08 & 94.4 & 0.33 & 0.08 & -0.006 & 0.09 & 94.2 & 0.33 & 0.08 \\ 
  A & C & 0.003 & 0.07 & 94.0 & 0.26 & 0.07 & -0.001 & 0.07 & 94.8 & 0.26 & 0.07 \\ 
  B1 & A & -0.012 & 0.07 & 92.8 & 0.26 & 0.07 & -0.011 & 0.07 & 92.4 & 0.26 & 0.07 \\ 
  B1 & B & -0.003 & 0.09 & 92.4 & 0.33 & 0.08 & -0.004 & 0.08 & 94.0 & 0.33 & 0.08 \\ 
  B1 & C & -0.012 & 0.07 & 93.2 & 0.26 & 0.07 & -0.011 & 0.07 & 93.6 & 0.26 & 0.07 \\ 
  B2 & A & -0.100 & 0.07 & 59.8 & 0.27 & 0.07 & -0.118 & 0.07 & 46.4 & 0.27 & 0.07 \\ 
  B2 & B & 0.004 & 0.09 & 91.4 & 0.33 & 0.08 & 0.001 & 0.09 & 92.8 & 0.33 & 0.08 \\ 
  B2 & C & -0.100 & 0.07 & 71.0 & 0.27 & 0.07 & -0.118 & 0.07 & 61.0 & 0.27 & 0.07 \\ 
  B3 & A & -0.024 & 0.07 & 91.0 & 0.26 & 0.07 & -0.028 & 0.07 & 92.0 & 0.27 & 0.07 \\ 
  B3 & B & 0.003 & 0.08 & 94.0 & 0.33 & 0.08 & 0.002 & 0.08 & 94.6 & 0.33 & 0.08 \\ 
  B3 & C & -0.025 & 0.07 & 91.6 & 0.26 & 0.07 & -0.028 & 0.07 & 93.8 & 0.27 & 0.07 \\ 
  B4 & A & -0.041 & 0.07 & 87.2 & 0.26 & 0.07 & -0.038 & 0.07 & 89.2 & 0.26 & 0.07 \\ 
  B4 & B & 0.001 & 0.08 & 93.8 & 0.33 & 0.08 & -0.000 & 0.09 & 92.4 & 0.33 & 0.08 \\ 
  B4 & C & -0.041 & 0.07 & 89.0 & 0.26 & 0.07 & -0.038 & 0.07 & 89.8 & 0.26 & 0.07 \\ 
   \hline
\end{tabular}}
\end{table}

Table \ref{table: concurrent_small_effect} and \ref{table: concurrent_medium_effect}  summarize the simulation results for the concurrent trial design with two different effect sizes. We have the following observations. First, when the true data generating process follows a constant treatment effect model, \textsf{Model A} and \textsf{C} both have favorable performances in that the bias is small, the coverage is close to the nominal level, and the standard error is small. While \textsf{Model B} is unbiased, it tends to be less efficient compared to Models A and C and often exhibits coverage rates slightly below the nominal level.

Second, when the data are generated from a time-varying fixed treatment effect model, fitting a model assuming a constant (\textsf{Model A}) or random normally distributed effect (\textsf{Model C}) could lead to a sometimes large bias. 
For example, when $T = 11$, $\Delta_1 = 0.29$ and $\Delta_2 = 0.40$, under \textsf{Outcome Model B3}, \textsf{Model A} yields a coverage rate of $72.4\%$ with a bias of $-0.07$, which further deteriorates to $0\%$ coverage and a bias of $-0.37$ under \textsf{Outcome Model B2}.
Compared to \textsf{Model A}, \textsf{Model C} often achieves closer-to-nominal coverage in most scenarios. This is because the bias-to-SE (standard error) ratio determines the estimator coverage, and \textsf{Model C} tends to have similar SE to \textsf{Model A}, while featuring smaller bias across many settings. However, fitting a random effect model (\textsf{Model C}) may also lead to sub-nominal coverage when the bias is large relative to the SE. This helps explain why the performance of \textsf{Model C} in our study differs from that reported by Maleyeff et al. (2022) \cite{maleyeff23}. In their simulations with a single intervention, the bias-to-SE ratio for \textsf{Model C} typically ranges from $5\%$ to $20\%$, yielding coverage rates above $90\%$. In contrast, our data generating process allows for a broader range of bias-to-SE ratios -- from as low as $6\%$ to as high as $360\%$ -- depending on the pattern of the time-varying effects. 
Accordingly, the coverage of \textsf{Model C}, when the true data generating process follows \textsf{Model B}, varies greatly from substantial  undercoverage (e.g., $59.9\%$ for Intervention 1 under \textsf{Outcome Model B4} when $T = 11$, $\Delta_1 = 0.29$, and $\Delta_2 = 0.40$) to near-nominal coverage (e.g., $95.6 \%$ for Intervention 1 under \textsf{Outcome Model B1} when $T = 5$, $\Delta_1 = 0.10$, and $\Delta_2 = 0.14$).

We first examine the bias incurred when using \textsf{Model C} to fit data generated under \textsf{Model B} with various exposure-time-specific effect patterns.
If the time-varying treatment effects follow an approximately normal pattern, the resulting bias tends to be minimal. For instance, when $T = 11$, $\Delta_1 = 0.29$, and $\Delta_2 = 0.40$ with a linear increasing effect, \textsf{Model C} achieves a coverage rate of $91.2\%$, for the average treatment effect of Intervention 1, with a bias of $-0.033$ and SE of $0.07$. 
However, when treatment effects follow lagged patterns, a random effect fit assuming a normal working model could incur substantial bias, and the magnitude of bias appears sensitive to the specific lag structure. For example, when $T = 11$, $\Delta_1 = 0.29$ and $\Delta_2 = 0.40$, the bias is $-0.07$ and coverage is $84.0\%$ under a one-period lag, whereas the bias increases to $-0.29$ and coverage drops to $2.4\%$ when the effect lags for half of the exposure duration. For both lagged patterns, the SE of the estimator remains largely the same and the deterioration in coverage is driven by the increase in bias.
Moreover, when the model is misspecified, increasing the number of periods (e.g., from $T = 5$ to $T = 11$) leads to worse coverage: as $T$ increases, bias persists while the variance of the estimator decreases as information accrues, pushing the bias-to-SE ratio higher. 

Notably, \textsf{Model B} consistently exhibits small bias across all effect patterns and achieves coverage close to the nominal level. These trends are also observed in the factorial design, as shown in Tables \ref{table: facrotial_small_effect} and \ref{table: facrotial_medium_effect}, where results are qualitatively similar to those in the concurrent design.

Overall, the performance of each model is  driven by the interplay between bias and variance, which in turn is shaped by the time-varying treatment effect patterns, effect sizes, and the number of exposure periods (i.e., effective sample size). These simulation results suggest that in the absence of prior knowledge about effect heterogeneity over time, using a time-varying fixed treatment effect model appears to be most desirable, as it has robust performance under both constant and time-varying treatment effect scenarios. On the other hand, if the researcher believes that the effects across time periods would follow a close-to-normal pattern, then fitting a random treatment effect model would have small bias and tend to be more efficient compared to the fixed treatment effect model.

\begin{table}[ht]
\centering
\caption{Comparison of model estimates for the exposure-time-averaged treatment effects under different types of treatment effect heterogeneity across exposure time in a concurrent design, with a large effect size and 
 $T=5$ or $T=11$. We use $\hat{\Delta}$ to denote the estimate of the exposure-time-averaged treatment effect, $\hat{\E}(\cdot)$ and $\hat{\text{SD}}(\cdot)$ to denote the sample mean and sample standard deviation across $500$ simulations, and $\hat{\text{SE}}(\cdot)$ to denote the estimated standard error.}
 \label{table: concurrent_medium_effect}
\resizebox{\textwidth}{!}{
\begin{tabular}{cccccccccccc}
  \hline
    & &\multicolumn{5}{c}{$\Delta_1$} & \multicolumn{5}{c}{$\Delta_2$} \\ \cline{3-7} \cline{8-12}
    Outcome & Fitting &    &  & CI  &  &  &   & &CI  &  \\ 
Model& Model & Bias &  $\hat{\text{SD}}(\hat{\Delta})$ &  Coverage (\%) & Length & $\hat{\E}(\hat{\text{SE}}(\hat{\Delta}))$  & Bias  & $\hat{\text{SD}}(\hat{\Delta})$ & Coverage (\%) & Length & $\hat{\E}(\hat{\text{SE}}(\hat{\Delta}))$ \\ 
  \hline
\multicolumn{12}{c}{\textbf{T=5, $\Delta_1=0.28$, $\Delta_2 = 0.40$}}\\
A & A & -0.006 & 0.16 & 96.0 & 0.66 & 0.17 & 0.004 & 0.16 & 94.2 & 0.66 & 0.17 \\ 
  A & B & -0.006 & 0.20 & 93.6 & 0.79 & 0.20 & 0.004 & 0.19 & 95.4 & 0.78 & 0.20 \\ 
  A & C & -0.007 & 0.17 & 95.8 & 0.69 & 0.18 & 0.005 & 0.16 & 94.8 & 0.68 & 0.18 \\ 
  B1 & A & -0.029 & 0.17 & 94.0 & 0.66 & 0.17 & -0.020 & 0.17 & 94.0 & 0.66 & 0.17 \\ 
  B1 & B & 0.004 & 0.20 & 94.2 & 0.79 & 0.20 & 0.020 & 0.20 & 94.0 & 0.79 & 0.20 \\ 
  B1 & C & -0.029 & 0.17 & 95.4 & 0.69 & 0.18 & -0.020 & 0.17 & 94.0 & 0.69 & 0.18 \\ 
  B2 & A & -0.328 & 0.18 & 41.9 & 0.72 & 0.19 & -0.388 & 0.19 & 30.5 & 0.74 & 0.19 \\ 
  B2 & B & -0.001 & 0.19 & 96.2 & 0.79 & 0.20 & -0.003 & 0.19 & 96.0 & 0.79 & 0.20 \\ 
  B2 & C & -0.275 & 0.18 & 63.1 & 0.72 & 0.18 & -0.329 & 0.18 & 53.1 & 0.72 & 0.19 \\ 
  B3 & A & -0.162 & 0.17 & 76.4 & 0.69 & 0.18 & -0.192 & 0.17 & 72.4 & 0.69 & 0.18 \\ 
  B3 & B & 0.017 & 0.19 & 94.8 & 0.79 & 0.20 & 0.015 & 0.19 & 95.6 & 0.79 & 0.20 \\ 
  B3 & C & -0.145 & 0.17 & 87.8 & 0.70 & 0.18 & -0.174 & 0.17 & 85.6 & 0.70 & 0.18 \\ 
  B4 & A & -0.142 & 0.16 & 83.8 & 0.68 & 0.17 & -0.139 & 0.16 & 82.6 & 0.68 & 0.17 \\ 
  B4 & B & 0.008 & 0.18 & 96.4 & 0.80 & 0.20 & 0.012 & 0.17 & 97.6 & 0.79 & 0.20 \\ 
  B4 & C & -0.138 & 0.16 & 89.6 & 0.70 & 0.18 & -0.136 & 0.16 & 88.0 & 0.70 & 0.18 \\ 
   \hline
   \multicolumn{12}{c}{\textbf{T=11, $\Delta_1=0.29$, $\Delta_2 = 0.40$}}\\
A & A & -0.006 & 0.07 & 95.4 & 0.26 & 0.07 & -0.006 & 0.07 & 94.0 & 0.26 & 0.07 \\ 
  A & B & -0.007 & 0.09 & 93.0 & 0.33 & 0.08 & -0.006 & 0.09 & 92.4 & 0.33 & 0.08 \\ 
  A & C & -0.006 & 0.07 & 95.4 & 0.26 & 0.07 & -0.007 & 0.07 & 94.2 & 0.26 & 0.07 \\ 
  B1 & A & -0.033 & 0.07 & 90.0 & 0.26 & 0.07 & -0.034 & 0.07 & 89.2 & 0.26 & 0.07 \\ 
  B1 & B & 0.000 & 0.09 & 95.4 & 0.33 & 0.08 & -0.000 & 0.09 & 93.2 & 0.33 & 0.08 \\ 
  B1 & C & -0.033 & 0.07 & 91.2 & 0.27 & 0.07 & -0.035 & 0.07 & 90.6 & 0.26 & 0.07 \\ 
  B2 & A & -0.368 & 0.08 & 0.0 & 0.31 & 0.08 & -0.429 & 0.08 & 0.0 & 0.32 & 0.08 \\ 
  B2 & B & 0.001 & 0.08 & 94.6 & 0.33 & 0.08 & -0.001 & 0.08 & 94.8 & 0.33 & 0.08 \\ 
  B2 & C & -0.289 & 0.08 & 2.4 & 0.29 & 0.08 & -0.339 & 0.08 & 0.4 & 0.30 & 0.08 \\ 
  B3 & A & -0.073 & 0.07 & 72.4 & 0.27 & 0.07 & -0.088 & 0.07 & 65.0 & 0.27 & 0.07 \\ 
  B3 & B & 0.010 & 0.09 & 92.4 & 0.33 & 0.08 & 0.007 & 0.09 & 92.6 & 0.33 & 0.08 \\ 
  B3 & C & -0.069 & 0.07 & 84.0 & 0.27 & 0.07 & -0.082 & 0.07 & 77.8 & 0.27 & 0.07 \\ 
  B4 & A & -0.123 & 0.07 & 43.9 & 0.27 & 0.07 & -0.115 & 0.07 & 48.7 & 0.27 & 0.07 \\ 
  B4 & B & 0.007 & 0.08 & 94.8 & 0.33 & 0.08 & 0.007 & 0.08 & 93.0 & 0.33 & 0.08 \\ 
  B4 & C & -0.122 & 0.07 & 59.9 & 0.27 & 0.07 & -0.114 & 0.07 & 64.5 & 0.27 & 0.07 \\ 
   \hline
\end{tabular}}
\end{table}

\begin{table}[ht]
\centering
\caption{Comparison of model estimates for the exposure-time-averaged treatment effects under different types of treatment effect heterogeneity across exposure time in a factorial design, with a small effect size and 
 $T=5$ or $T=11$. We use $\hat{\Delta}$ to denote the estimate of the exposure-time-averaged treatment effect, $\hat{\E}(\cdot)$ and $\hat{\text{SD}}(\cdot)$ to denote the sample mean and sample standard deviation across $500$ simulations, and $\hat{\text{SE}}(\cdot)$ to denote the estimated standard error.} 
\label{table: facrotial_small_effect}
\resizebox{\textwidth}{!}{
\begin{tabular}{cccccccccccc}
  \hline
    & &\multicolumn{5}{c}{$\Delta_1$} & \multicolumn{5}{c}{$\Delta_2$} \\ \cline{3-7} \cline{8-12}
    Outcome & Fitting &    &  & CI  &  &  &   & &CI  &  \\ 
 Model& Model & Bias &  $\hat{\text{SD}}(\hat{\Delta})$ &  Coverage (\%) & Length & $\hat{\E}(\hat{\text{SE}}(\hat{\Delta}))$  & Bias  & $\hat{\text{SD}}(\hat{\Delta})$ & Coverage (\%) & Length & $\hat{\E}(\hat{\text{SE}}(\hat{\Delta}))$ \\ 
  \hline
\multicolumn{12}{c}{\textbf{T=5, $\Delta_1=0.10 $, $\Delta_2 = 0.14$}}\\
A & A & -0.003 & 0.14 & 94.8 & 0.57 & 0.15 & 0.010 & 0.15 & 93.8 & 0.57 & 0.15 \\ 
  A & B & -0.003 & 0.21 & 95.2 & 0.87 & 0.22 & 0.012 & 0.22 & 96.0 & 0.87 & 0.22 \\ 
  A & C & -0.004 & 0.14 & 96.0 & 0.61 & 0.16 & 0.011 & 0.15 & 95.0 & 0.61 & 0.16 \\ 
  B1 & A & -0.001 & 0.14 & 95.0 & 0.57 & 0.15 & -0.014 & 0.14 & 95.2 & 0.57 & 0.15 \\ 
  B1 & B & 0.003 & 0.21 & 94.4 & 0.88 & 0.23 & -0.002 & 0.21 & 96.0 & 0.88 & 0.22 \\ 
  B1 & C & -0.002 & 0.14 & 96.2 & 0.62 & 0.16 & -0.013 & 0.14 & 95.8 & 0.61 & 0.16 \\ 
  B2 & A & -0.070 & 0.14 & 91.0 & 0.58 & 0.15 & -0.098 & 0.15 & 85.8 & 0.58 & 0.15 \\ 
  B2 & B & -0.019 & 0.21 & 95.4 & 0.88 & 0.23 & 0.011 & 0.22 & 95.6 & 0.87 & 0.22 \\ 
  B2 & C & -0.069 & 0.14 & 92.8 & 0.63 & 0.16 & -0.100 & 0.15 & 90.2 & 0.63 & 0.16 \\ 
   B3 & A & 0.012 & 0.15 & 94.0 & 0.57 & 0.15 & -0.036 & 0.14 & 94.8 & 0.57 & 0.15 \\ 
  B3 & B & 0.006 & 0.22 & 95.0 & 0.88 & 0.23 & -0.001 & 0.22 & 95.2 & 0.89 & 0.23 \\ 
  B3 & C & 0.008 & 0.15 & 95.2 & 0.64 & 0.16 & -0.042 & 0.14 & 96.6 & 0.64 & 0.16 \\ 
  B4 & A & -0.036 & 0.15 & 94.2 & 0.57 & 0.15 & -0.011 & 0.14 & 94.2 & 0.57 & 0.15 \\ 
  B4 & B & 0.011 & 0.22 & 94.0 & 0.88 & 0.23 & 0.002 & 0.21 & 95.6 & 0.87 & 0.22 \\ 
  B4 & C & -0.038 & 0.15 & 95.8 & 0.62 & 0.16 & -0.011 & 0.14 & 96.0 & 0.61 & 0.16 \\ 
   \hline
     \multicolumn{12}{c}{\textbf{T=11, $\Delta_1=0.10$, $\Delta_2 = 0.13$}}\\
A & A & 0.000 & 0.06 & 93.6 & 0.23 & 0.06 & -0.002 & 0.06 & 94.4 & 0.23 & 0.06 \\ 
  A & B & -0.004 & 0.09 & 94.8 & 0.35 & 0.09 & -0.003 & 0.09 & 95.4 & 0.35 & 0.09 \\ 
  A & C & 0.000 & 0.06 & 94.4 & 0.23 & 0.06 & -0.001 & 0.06 & 94.6 & 0.23 & 0.06 \\ 
  B1 & A & -0.011 & 0.06 & 94.6 & 0.23 & 0.06 & -0.003 & 0.06 & 95.8 & 0.23 & 0.06 \\ 
  B1 & B & -0.005 & 0.10 & 93.0 & 0.35 & 0.09 & 0.003 & 0.10 & 92.6 & 0.35 & 0.09 \\ 
  B1 & C & -0.011 & 0.06 & 94.2 & 0.23 & 0.06 & -0.003 & 0.06 & 97.4 & 0.23 & 0.06 \\ 
  B2 & A & -0.054 & 0.06 & 81.4 & 0.23 & 0.06 & -0.087 & 0.06 & 58.3 & 0.23 & 0.06 \\ 
  B2 & B & 0.002 & 0.09 & 94.4 & 0.36 & 0.09 & 0.005 & 0.10 & 93.8 & 0.36 & 0.09 \\ 
  B2 & C & -0.054 & 0.06 & 84.8 & 0.24 & 0.06 & -0.087 & 0.06 & 66.5 & 0.24 & 0.06 \\ 
 B3 & A & -0.011 & 0.06 & 92.4 & 0.23 & 0.06 & -0.021 & 0.06 & 89.6 & 0.23 & 0.06 \\ 
  B3 & B & 0.001 & 0.09 & 93.0 & 0.35 & 0.09 & -0.002 & 0.10 & 92.4 & 0.35 & 0.09 \\ 
  B3 & C & -0.011 & 0.06 & 93.6 & 0.23 & 0.06 & -0.021 & 0.06 & 91.4 & 0.23 & 0.06 \\ 
  B4 & A & -0.041 & 0.06 & 87.2 & 0.23 & 0.06 & -0.005 & 0.06 & 94.2 & 0.23 & 0.06 \\ 
  B4 & B & -0.000 & 0.09 & 93.8 & 0.35 & 0.09 & 0.003 & 0.09 & 94.2 & 0.36 & 0.09 \\ 
  B4 & C & -0.042 & 0.06 & 88.6 & 0.23 & 0.06 & -0.005 & 0.06 & 93.6 & 0.23 & 0.06 \\ 
   \hline
\end{tabular}}
\end{table}

\begin{table}[ht]
\centering
\caption{Comparison of model estimates for the exposure-time-averaged treatment effects under different types of treatment effect heterogeneity across exposure time in a factorial design, with a large effect size and 
 $T=5$ or $T=11$. We use $\hat{\Delta}$ to denote the estimate of the exposure-time-averaged treatment effect, $\hat{\E}(\cdot)$ and $\hat{\text{SD}}(\cdot)$ to denote the sample mean and sample standard deviation across $500$ simulations, and $\hat{\text{SE}}(\cdot)$ to denote the estimated standard error.} 
 \label{table: facrotial_medium_effect}
\resizebox{\textwidth}{!}{
\begin{tabular}{cccccccccccc}
  \hline
    & &\multicolumn{5}{c}{$\Delta_1$} & \multicolumn{5}{c}{$\Delta_2$} \\ \cline{3-7} \cline{8-12}
    Outcome & Fitting &    &  &   & CI &  &   & &  &CI  \\ 
 Model& Model & Bias &  $\hat{\text{SD}}(\hat{\Delta})$ &  Coverage (\%) & Length & $\hat{\E}(\hat{\text{SE}}(\hat{\Delta}))$  & Bias  & $\hat{\text{SD}}(\hat{\Delta})$ & Coverage (\%) & Length & $\hat{\E}(\hat{\text{SE}}(\hat{\Delta}))$ \\ 
  \hline
\multicolumn{12}{c}{\textbf{T=5, $\Delta_1=0.28$, $\Delta_2 = 0.40$}}\\
A & A & 0.001 & 0.14 & 95.0 & 0.57 & 0.15 & 0.006 & 0.14 & 95.6 & 0.57 & 0.15 \\ 
  A & B & 0.000 & 0.22 & 94.6 & 0.88 & 0.23 & 0.004 & 0.21 & 95.6 & 0.88 & 0.23 \\ 
  A & C & 0.001 & 0.14 & 94.8 & 0.61 & 0.16 & 0.006 & 0.14 & 96.0 & 0.61 & 0.16 \\ 
  B1 & A & -0.016 & 0.14 & 94.4 & 0.57 & 0.15 & -0.022 & 0.15 & 93.8 & 0.57 & 0.15 \\ 
  B1 & B & 0.011 & 0.21 & 95.2 & 0.87 & 0.22 & -0.011 & 0.22 & 93.8 & 0.87 & 0.22 \\ 
  B1 & C & -0.018 & 0.14 & 95.8 & 0.61 & 0.16 & -0.024 & 0.15 & 95.0 & 0.62 & 0.16 \\ 
  B2 & A & -0.218 & 0.16 & 63.3 & 0.63 & 0.16 & -0.338 & 0.17 & 32.1 & 0.65 & 0.17 \\ 
  B2 & B & -0.018 & 0.22 & 94.8 & 0.87 & 0.22 & 0.019 & 0.21 & 95.6 & 0.88 & 0.23 \\ 
  B2 & C & -0.174 & 0.19 & 85.6 & 0.74 & 0.19 & -0.281 & 0.18 & 66.5 & 0.75 & 0.19 \\ 
  B3 & A & 0.006 & 0.15 & 94.0 & 0.58 & 0.15 & -0.117 & 0.14 & 83.8 & 0.59 & 0.15 \\ 
  B3 & B & -0.001 & 0.22 & 95.6 & 0.88 & 0.23 & -0.014 & 0.21 & 94.4 & 0.89 & 0.23 \\ 
  B3 & C & -0.009 & 0.18 & 98.0 & 0.78 & 0.20 & -0.135 & 0.17 & 90.0 & 0.78 & 0.20 \\ 
  B4 & A & -0.124 & 0.15 & 81.0 & 0.59 & 0.15 & -0.053 & 0.15 & 89.0 & 0.58 & 0.15 \\ 
  B4 & B & 0.003 & 0.23 & 93.8 & 0.87 & 0.22 & -0.007 & 0.22 & 94.4 & 0.88 & 0.23 \\ 
  B4 & C & -0.127 & 0.15 & 85.6 & 0.65 & 0.17 & -0.059 & 0.16 & 93.0 & 0.65 & 0.17 \\ 
   \hline
     \multicolumn{12}{c}{\textbf{T=11, $\Delta_1=0.29$, $\Delta_2 = 0.40$}}\\
A & A & 0.008 & 0.06 & 93.8 & 0.23 & 0.06 & -0.002 & 0.06 & 94.2 & 0.23 & 0.06 \\ 
  A & B & 0.005 & 0.10 & 92.0 & 0.36 & 0.09 & 0.002 & 0.09 & 95.0 & 0.36 & 0.09 \\ 
  A & C & 0.008 & 0.06 & 93.6 & 0.23 & 0.06 & -0.002 & 0.06 & 94.8 & 0.23 & 0.06 \\ 
  B1 & A & -0.021 & 0.06 & 92.8 & 0.23 & 0.06 & -0.017 & 0.06 & 93.0 & 0.23 & 0.06 \\ 
  B1 & B & -0.000 & 0.09 & 93.6 & 0.36 & 0.09 & 0.003 & 0.09 & 95.8 & 0.36 & 0.09 \\ 
  B1 & C & -0.022 & 0.06 & 94.0 & 0.23 & 0.06 & -0.017 & 0.06 & 93.6 & 0.23 & 0.06 \\ 
  B2 & A & -0.200 & 0.07 & 5.6 & 0.26 & 0.07 & -0.327 & 0.07 & 0.0 & 0.27 & 0.07 \\ 
  B2 & B & -0.004 & 0.09 & 93.4 & 0.35 & 0.09 & 0.003 & 0.09 & 92.2 & 0.35 & 0.09 \\ 
  B2 & C & -0.158 & 0.08 & 29.3 & 0.28 & 0.07 & -0.273 & 0.09 & 1.6 & 0.28 & 0.07 \\ 
  B3 & A & -0.030 & 0.06 & 86.2 & 0.23 & 0.06 & -0.064 & 0.06 & 72.8 & 0.23 & 0.06 \\ 
  B3 & B & -0.000 & 0.09 & 94.0 & 0.35 & 0.09 & -0.002 & 0.09 & 93.6 & 0.35 & 0.09 \\ 
  B3 & C & -0.033 & 0.06 & 89.6 & 0.24 & 0.06 & -0.059 & 0.06 & 80.0 & 0.24 & 0.06 \\ 
  B4 & A & -0.122 & 0.06 & 34.1 & 0.23 & 0.06 & -0.036 & 0.06 & 86.0 & 0.23 & 0.06 \\ 
  B4 & B & -0.002 & 0.09 & 92.8 & 0.35 & 0.09 & 0.004 & 0.10 & 94.6 & 0.36 & 0.09 \\ 
  B4 & C & -0.122 & 0.06 & 42.5 & 0.24 & 0.06 & -0.037 & 0.06 & 88.8 & 0.24 & 0.06 \\ 
   \hline
\end{tabular}}
\end{table}

\subsection{Comparing the empirical characteristics under concurrent and factorial designs}\label{subsec: simu compare designs}
\subsubsection{Aims}

The aim of this simulation was to compare the empirical power of the concurrent and factorial designs in detecting the exposure-time-averaged treatment effects of multiple interventions in the presence of time-varying treatment effects. 

\subsubsection{Data-generating mechanisms}
Each design consisted of 5 time periods, 8 clusters, and $n$ individuals sampled cross-sectionally from each cluster at every time point. We considered three cluster-period sample sizes:  $n=\{30, 100, 500\}$. The configuration of the concurrent design is shown in Figure \ref{fig:concurrent}. To facilitate a fair comparison with the same number of clusters, the factorial design configuration depicted in Figure \ref{fig:factorial design} was augmented by adding two clusters: one transitioning from control to Intervention 1, and the other from control to Intervention 2, in the final time period. 

Time-varying treatment effects were modeled as linearly increasing, as specified in \textsf{Outcome Model B1} in Section \ref{Sim 1: DGP}, with $ \alpha_i\sim \N(0,0.05)$ and $ \epsilon_{ijs}\sim \N(0,0.95)$. For both designs, we evaluated statistical power across a range of positive exposure-time-averaged effect sizes: $\Delta_1 =\{0.01,0.11,0.21,…,0.61\}$ in increments of 0.1 for Intervention 1, and $\Delta_2= \Delta_1 + 0.28$ for Intervention 2. 

\subsubsection{Estimands}
As in the simulation study in Section \ref{subsec: simu compare estimators}, we target the exposure-time-averaged treatment effects for each intervention, as defined in equation~\eqref{causal estimand}.

\subsubsection{Methods}
As indicated by the simulation results in Section \ref{sim1: results}, the Type I error rates from the random treatment effect model did not consistently meet the nominal level. 
Therefore, we based our power analyses on the time-varying fixed treatment effect model defined in equation~\eqref{model: time-varying fixed}. 
For each data generating process, we conducted 500 simulations. Confidence intervals for the exposure-time-averaged treatment effects were constructed using 500 bootstrap samples within each cluster-period.

\subsubsection{Performance measures}
The primary performance metric is empirical power, defined as the proportion of simulations in which the null hypothesis of no average effect is rejected at the $5\%$ significance level.

\begin{figure}[ht]
    \centering
    \includegraphics[width=0.65\linewidth]{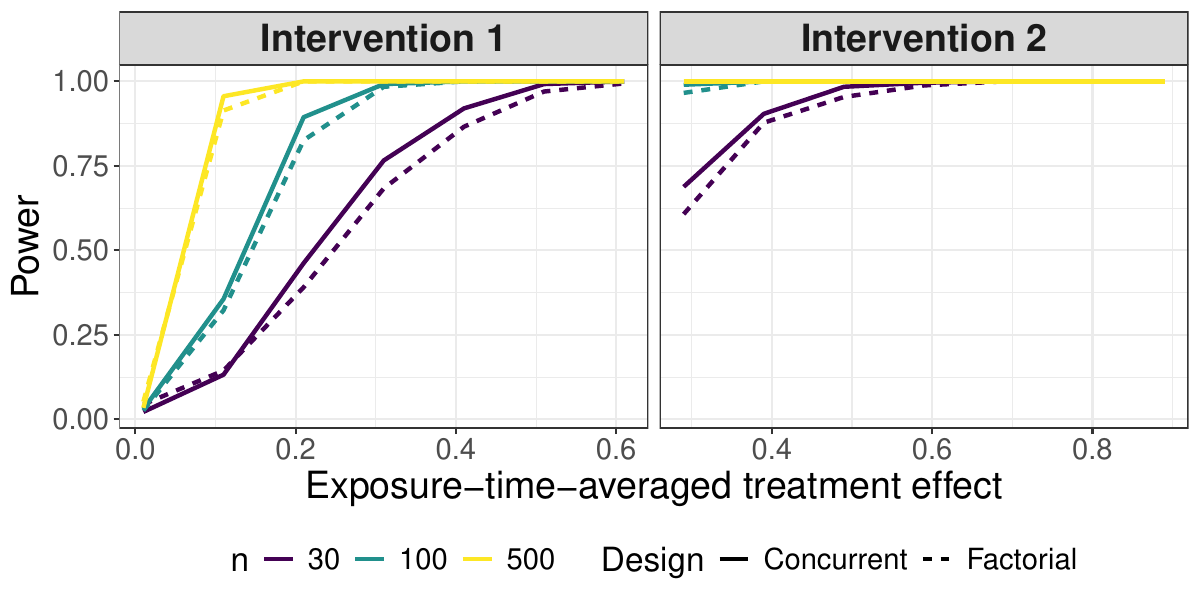}
    \caption{Comparison of power for detecting exposure-time-averaged treatment effects between a concurrent design and a factorial design, based on the time-varying fixed treatment effect model. }
    \label{fig:power based on model B}
\end{figure}

Figure \ref{fig:power based on model B} displays the power of detecting the exposure-time-averaged treatment effect of each intervention under both concurrent and factorial designs, based on the time-varying fixed treatment effect model. As expected, power increases with larger effect sizes and greater sample sizes.
For instance, when the exposure-time-averaged effect size for Intervention 1 is $0.2$, the concurrent design attains an approximately $90\%$ power with a sample size of $n=100$ per cluster. However, with a smaller sample size of $n=30$, the effect size must be as large as $0.4$ to achieve similar power.
In most settings, the concurrent design  achieves higher power than the factorial design, though the difference between the two designs remains relatively small, typically within $5\%$ to $10\%$. However, when the sample size is small (e.g., 
$n=30$) and the effect size is as low as $0.1$, the factorial design may achieve slightly higher power.
In clinical trials, researchers need to further weigh practical resource and domain knowledge constraints when choosing the most appropriate design, and our exploration provides an initial assessment of relative power under these two designs in the presence of time-varying treatment effects.

\section{An Illustrative Data Example: The PONDER-ICU Stepped Wedge Trial}
\label{sec: data analysis}
\begin{figure}[h]
    \centering
    \includegraphics[width=1\linewidth]{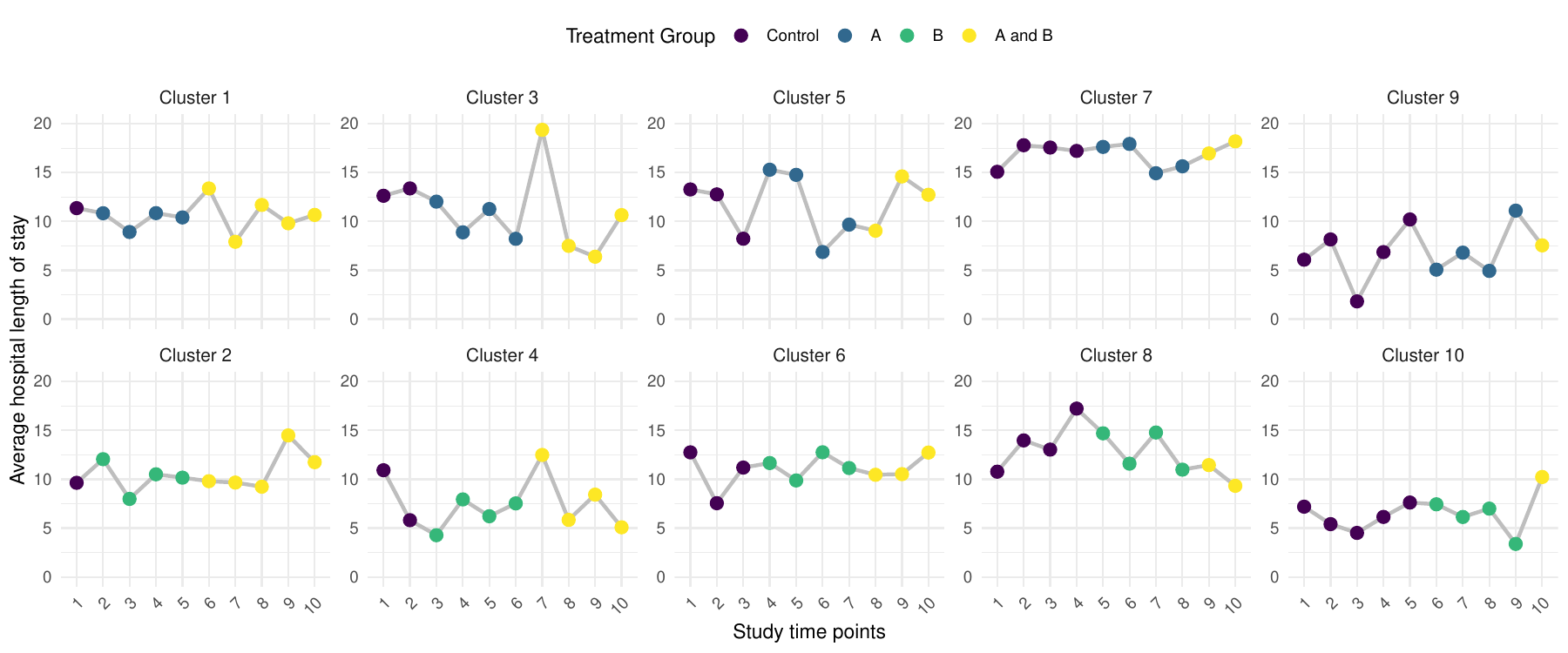}
    \caption{Average hospital
length of stay over time by cluster and treatment group.}
    \label{fig:ponder}
\end{figure}

To illustrate the application of the models described in Section \ref{model}, we conducted analyses using the Prognosticating Outcomes and Nudging Decisions in the Electronic health Record in the Intensive Care Unit (PONDER-ICU) dataset, 
which originates from a pragmatic, stepped-wedge, cluster-randomized trial conducted across 17 intensive care units (ICUs) in 10 hospitals within the Atrium Health system from February 2018 to October 2020. This trial aimed to evaluate the effectiveness of two clinician-targeted behavioral interventions embedded in the electronic health record to improve serious illness communication and end-of-life care processes for mechanically ventilated patients at high risk of mortality or severe functional morbidity. Both interventions were designed using behavioral economic principles for effective clinician nudges: (1) \textsf{Intervention A}: documenting six-month functional prognosis and (2) \textsf{Intervention B}: indicating whether a comfort-focused treatment alternative was offered, along with a justification if it was not. These interventions, either alone or combined, were compared to usual care to assess their impact on hospital length of stay and various secondary outcomes. The dataset includes 3500 ICU encounters among 3250 individuals, 10 clusters, and 10 time points. In our analysis, we focus on the primary outcome, hospital length of stay (LOS), at the individual patient-level. Figure \ref{fig:ponder} presents the average hospital LOS over time by cluster and treatment group.

We fit three models to the data to evaluate the exposure-time-averaged treatment effect for each intervention: a constant treatment effect model, a time-varying fixed treatment effect model, and a random treatment effect model. 
We restricted our analysis to the case where the joint effect of \textsf{Interventions A} and \textsf{B} equals the sum of their individual effects. This assumption is necessary because time-varying joint effects are not identifiable under a fixed treatment effect model.
Pointwise confidence intervals were obtained using bootstrapping of individual-level data within each cluster period.

Across all three models, the results consistently show no evidence of significant average treatment effects for either intervention. 
The estimated exposure-time-averaged treatment effects from the constant treatment effect model and the random treatment effect model were similar.
Specifically, from the constant treatment effect model, the estimated average treatment effect was $0.22~ ( -1.66, 2.09)$ for \textsf{Intervention A} and $0.30~ (-1.56, 2.32)$ for \textsf{Intervention B}. Similarly, the random treatment effect model produced estimates of  $0.21~   (-1.90, 1.58)$ for \textsf{Intervention A} and $0.33 ~(-2.02, 1.72)$ for \textsf{Intervention B}.
In contrast, the time-varying fixed effect model yielded noticeably wider confidence intervals: $-1.50~  (-4.23, 1.02)$ for \textsf{Intervention A} and $-1.73 ~ (-4.58, 1.20)$ for \textsf{Intervention B}. 
We repeated the analysis for the secondary outcome, time to comfort-care orders, and observed similar patterns. The constant treatment effect model estimated an average treatment effect of $0.19~(-1.78, 1.94)$ for \textsf{Intervention A} and $0.43~(-1.55, 2.56)$ for \textsf{Intervention B}. The random treatment effect model gave comparable estimates of $0.18~(-2.00, 1.55)$ and $0.45~(-1.98, 1.86)$, respectively. Once again, the time-varying fixed effect model produced estimates in the opposite direction, with wider confidence intervals: $-1.55~(-4.63, 0.98)$ for \textsf{Intervention A} and $-1.70~(-4.61, 1.38)$ for \textsf{Intervention B}.

To reconcile these findings, one plausible explanation is that the treatment effect is genuinely constant. In such cases, both the constant and random treatment effect models tend to yield more stable and consistent estimates, while the time-varying fixed effect model may introduce additional variability, resulting in less precise estimates. Similar behavior has been observed in simulation studies when data are generated under a constant treatment effect. 
This interpretation also appears reasonable when considering Figure \ref{fig:ponder}, which shows no discernible temporal trends in outcomes within or across intervention groups, and the average differences in outcomes between intervention groups are marginal relative to the overall mean hospital LOS. 

An alternative explanation is that the treatment effect truly varies over time. In this scenario, the constant and random treatment effect models could be biased due to model misspecification, whereas the time-varying fixed effect model would better capture the underlying effect structure. 
 The opposing signs of the estimates from the constant and random effect models, compared to those from the time-varying fixed effect model, could indicate such bias. However, when multiple sources of bias are present, it is unlikely that the constant and random effect models would produce estimates that align closely in both direction and magnitude. This makes it challenging to attribute the observed discrepancies solely to time-varying effects.

\section{Summary and discussion}\label{summary}

This paper makes several important methodological contributions to the analysis of complex stepped wedge designs. First, we develop and formalize a set of model extensions that accommodate the intricacies of multiple-intervention SWDs with time-varying and exposure-dependent treatment effects. These extensions are especially relevant in real-world implementation settings, where interventions may not exert constant effects over time and where exposure-time effect heterogeneity is plausible. 
Second, to our knowledge, this is the first study to explicitly demonstrate the bias induced by ignoring exposure-time effect heterogeneity in the context of multiple-intervention SWDs. 
 Our findings reveal that the generalized least squares estimator derived from the constant treatment effect model exhibits distinct behavior depending on the nature of the true treatment effects. When the treatment effects are indeed constant over time, the estimator is unbiased and recovers the true effects. However, when the treatment effects vary over time, the constant effect estimator produces bias that can be substantial and directionally misleading. This bias is influenced by several factors, including the variance components, design parameters such as the number of periods and the number of interventions, as well as the specific pattern of effect variation over exposure time. 
 Compared to traditional single-intervention SWDs, we find that bias patterns in multiple-intervention designs are more complex: the direction of bias depends not only on the shape of the true time-varying treatment effect but also on the design structure. For instance, under the same underlying exposure-time effect heterogeneity pattern, the estimated treatment effects can have opposite directions in concurrent and factorial designs. Moreover, the bias for a single intervention may depend not only on its own effect trajectory but also on the temporal structure of co-occurring interventions. These findings underscore the importance of evaluating exposure-time effect heterogeneity in the full design context, particularly when multiple interventions interact over time.

Our findings have direct implications for modeling strategies in practice. While it is tempting to begin with the simplest constant treatment effect model, we recommend engaging time-varying models as part of routine sensitivity analyses. The degree to which exposure-time effect heterogeneity manifests depends on the intervention mechanism, study duration, and contextual implementation factors. Time-varying fixed treatment effect model and random treatment effect model formulations each carry distinct strengths and limitations in this setting. Time-varying fixed treatment effect models are robust to different specifications of the effect curve over time but may suffer from instability with large confidence interval widths when effective sample size is small. Random treatment effect models may be more efficient, especially in trials with a large number of periods, but their performance is sensitive to the shape of the true effect trajectory. In our case study of the PONDER-ICU stepped wedge trial, the time-varying fixed treatment effect model yielded negative treatment effect estimates, whereas the random treatment effect model results closely resembled the constant effect model. This contrast suggests that when results differ substantially across models, the discrepancy may stem more from inefficiency or instability in fixed treatment effect model estimation rather than evidence for time-varying effects.
We emphasize that formal testing for exposure-time effect heterogeneity should not be used to select the primary analysis model. Model selection based on post hoc testing can inflate the Type I error rate and undermines the principle of pre-specification that underlies valid inference in trial analysis. Instead, such tests should be framed as tools to aid interpretation and to assess robustness across model choices. In trial planning, it is often impossible to predict whether time-varying effects will arise, and thus, a prespecified primary analysis model should be chosen based on prior knowledge and feasibility, supplemented by sensitivity analyses under alternative model forms.

There are several limitations and future directions. In this work, we have focused primarily on models with a simple cluster-level random intercept. However, the correlation structure in SWDs may be more complex in practice. Alternative specifications, such as nested exchangeable or exponential decay correlation models, could more accurately capture within-cluster dependencies. As reviewed in Li et al.\cite{overview}, these models have been widely studied, and it is important to acknowledge that our results can be generalized to settings where such complex correlation structures are present, but more extensive algebraic or simulation work (if closed-form expressions are unavailable) may be needed. Nevertheless, inference remains valid under correlation misspecification when using cluster-robust sandwich variance estimators\cite{ouyang2024maintaining}.
Additionally, our development has focused on continuous outcomes. Extensions to binary or count outcomes — especially under marginal models such as generalized estimating equations — remain a fertile area for future methodological work. Another direction involves incorporating covariate adjustment in the presence of exposure-time effect heterogeneity, which could yield more efficient estimators. While the model specification principles presented here can be extended to these settings, the corresponding bias formulas and variance estimators require careful adaptation. 
Future work should also address sample size and power calculations under these complex models. Maleyeff et al.\cite{maleyeff23} developed a sample size calculation framework for single-intervention SWDs under time-varying treatment effects using fixed treatment effect models. 
Sample size computations for multiple-intervention SWDs must account not only for traditional design resources such as the number of clusters, number of periods, and maximum exposure duration, but also for design-specific structural features, such as whether the interventions are rolled out concurrently, factorially, or sequentially.
Finally, while we have focused on exposure-time effect heterogeneity, more intricate forms of heterogeneity — such as those involving calendar time or combinations of calendar and exposure time (i.e., saturated effect models) — are highly relevant in pragmatic trials. These models further complicate identification and estimation but are essential for capturing implementation realities. We refer readers to recent developments in this area, including Wang et al.\cite{wang2024achieve}, 
Lee et al.\cite{lee2024analysis}, and Chen and Li\cite{chen2025model}, for ongoing efforts in modeling treatment effect heterogeneity along calendar time in SWDs.

\section*{Acknowledgements}
Research in this article was partially supported by the Patient-Centered Outcomes Research Institute\textsuperscript{\textregistered} (PCORI\textsuperscript{\textregistered} Awards ME-2020C1-19220 to M.O.H. and ME-2022C2-27676 to F.L). F.L and M.O.H are funded by the United States National Institutes of Health (NIH), National Heart, Lung, and Blood Institute (grant number R01-HL168202). All statements in this report, including its findings and conclusions, are solely those of the authors and do not necessarily represent the views of the NIH or PCORI\textsuperscript{\textregistered} or its Board of Governors or Methodology Committee.

\section*{Conflict of interest}

MOH has received consulting fees from Elsevier, the American Thoracic Society, and Unlearn.AI, all for work unrelated to the topics in this manuscript.

\section*{Supporting information}

The following supporting information is available as part of the online article: proofs of theorems, propositions and corollaries. 
R code is publicly available at \url{https://github.com/Zhe-Chen-1999/SWCRT-TEH-multiple-interventions}.

\bibliography{ref}

\end{document}